\documentclass[reprint,aps,prb,amsmath,amssymb,longbibliography, notitlepage]{revtex4-2}
\usepackage{bm, graphicx}
\usepackage{physics}
\usepackage{multirow}
\usepackage{tabularx}
\usepackage[colorlinks, linkcolor= blue, citecolor = blue, urlcolor=blue]{hyperref}
\usepackage{pifont}
\usepackage{cancel}
\usepackage{comment}
\usepackage{mathrsfs}

\def\nn{\nonumber}
\def\bea{\begin{eqnarray}}
\def\eea{\end{eqnarray}}
\def\be{\begin{equation}}
\def\ee{\end{equation}}

\def\e{\varepsilon}
\def\m{\mathcal}
\def\kb{{\bm k}}
\def\yes{\ding{51}}
\def\no{\ding{55}}

\newcolumntype{P}[1]{>{\centering\arraybackslash}p{#1}}
\begin{document}
\title{Asymmetric Scattering Drives Large Nonlinear Nernst and Seebeck Effects}

\author{Harsh Varshney}
\email{hvarshny@iitk.ac.in}
\affiliation{Department of Physics, Indian Institute of Technology, Kanpur-208016, India.}
\author{Amit Agarwal}
\email{amitag@iitk.ac.in}
\affiliation{Department of Physics, Indian Institute of Technology, Kanpur-208016, India.}
\begin{abstract}
The nonlinear Nernst and Seebeck effects (NNE and NSE) offer promising routes for thermoelectric energy conversion in non-magnetic systems. While intrinsic mechanisms such as the nonlinear Drude and Berry-curvature–dipole terms are well established, extrinsic contributions to thermoelectric responses arising from disorder-induced asymmetric scattering remain comparatively less explored, despite growing experimental evidence of their dominance. Here, we develop a unified semiclassical theory of NNE and NSE that incorporates skew scattering and side-jump processes, identifying four distinct extrinsic contributions to NNE and two for NSE. A systematic symmetry analysis shows that these responses are allowed in time-reversal–symmetric non-magnets, $\mathcal{PT}$-symmetric antiferromagnets, and non-centrosymmetric magnetic systems such as altermagnets. As a case study, we demonstrate that ABA-stacked trilayer graphene hosts large nonlinear Nernst and Seebeck responses dominated by extrinsic scattering, in excellent agreement with recent experiments. Our results establish the microscopic origin of these effects and provide guiding principles for designing high-efficiency nonlinear thermoelectric devices.
\end{abstract}
\maketitle

\section{Introduction}

The Nernst and Seebeck effects generate transverse and longitudinal voltages under an applied temperature gradient~\cite{uchida2008observation, checklesky2009thermpower, kamran2016nernst, ikhlas2017large, Rana2018thermopower, kamal2019berry, Sakai2020Nature, PhysRevResearch.2.013088, zhou2022fundamental, Uchida2022thermoelectrics, Pasquale2024, PhysRevB.102.205414}. Their ability to convert heat into electrical energy underpins applications ranging from industrial waste-heat recovery to cryogenic space instrumentation~\cite{disalvo1999thermoelectric, Snyder2009thermoelectric, tan2011sustainable, koumoto2013thermoelectric, sothmann2014thermoelectric, chhatrasal2016, amin2020review, takao2021, papaj2021enhanced, yang2023flexible, yu2024ambient, wan2025extrinsic}. Recent efforts have exploited the anomalous Nernst effect in magnetic materials to achieve magnetic-field–free thermoelectric operation~\cite{ikhlas2017large, Sakai2018giant, guin2019anomalous, Yang2020giant, Uchida2021transverse, Pan2021giant, Li2023large}. However, stabilizing or aligning magnetization typically requires external fields, posing a fundamental obstacle for integration into low-dimensional quantum devices. 

\begin{figure}[t!]
    \centering
    \includegraphics[width=\linewidth]{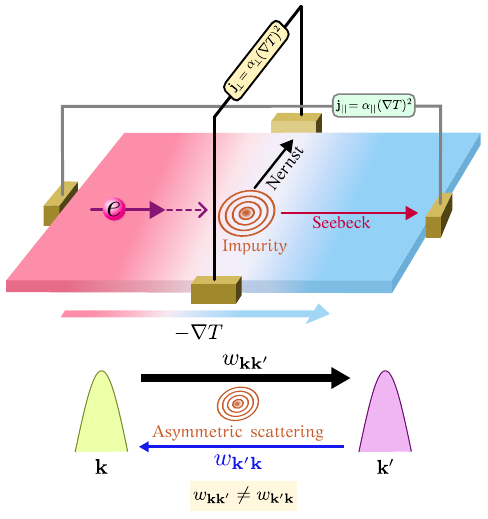}
    \caption{{\bf Schematic illustration of extrinsic nonlinear Nernst and Seebeck effects.} The upper panel depicts the NNE and NSE scaling quadratically with the applied temperature gradient [$\propto~(\nabla T)^2$]. Under a thermal bias, Bloch electrons scatter asymmetrically from $\kb$ to $\kb'$ due to impurities (lower panel), generating transverse (Nernst) and longitudinal (Seebeck) nonlinear currents. For clarity, the conventional linear Seebeck current is not shown explicitly.}
    \label{fig1}
\end{figure}

These limitations have motivated the search for non-magnetic alternatives such as graphene multilayers and other van der Waals materials. In these platforms, time-reversal symmetry forbids the linear Nernst effect. As a result, the nonlinear Nernst and Seebeck effects (NNE and NSE $\propto (\nabla T)^2$, see Fig.~\ref{fig1}) become the leading-order thermoelectric responses in non-magnetic systems~\cite{yu2019topological, marchegiani2020nonlinear, Chakraborty_2022, Wu2021nonlinear, Arisawa2024observation, liu2025nonlinear, hirata2025nonlinear, varshney2025intrinsic}. At the microscopic level, these responses arise from two broad classes of processes: intrinsic and symmetric-scattering mediated contributions such as the nonlinear Drude and Berry-curvature–dipole terms~\cite{zeng2019nonlinear, yu2019topological, Wu2021nonlinear, varshney2025intrinsic, varshney2023quantum, varshney2025intrinsic, Adak_2024}, and extrinsic mechanisms originating from disorder-induced asymmetric scattering, including side-jump and skew-scattering processes~\cite{du2019disorder,Sinha_2023, Datta2024nonlinear, Ahmed2025detecting, zhou2022fundamental}. While intrinsic nonlinear mechanisms are now well established, a unified framework that systematically incorporates all asymmetric-scattering channels remains lacking. More importantly, the asymmetric-scattering contributions can parametrically dominate the nonlinear thermoelectric responses, as highlighted by the recent discovery of a giant NNE in ABA-stacked trilayer graphene~\cite{liu2025nonlinear}. 

Here, we develop a unified semiclassical kinetic theory of the nonlinear Nernst and Seebeck effects that systematically incorporates disorder-mediated asymmetric scattering contributions beyond the symmetric-scattering approximation. Extending the Boltzmann formalism, we identify six distinct second-order thermoelectric response channels, comprising the nonlinear Drude and Berry-curvature–driven contributions together with four additional extrinsic channels arising from side-jump and skew-scattering processes. These extrinsic mechanisms generate qualitatively distinct contributions to both the nonlinear Nernst and Seebeck effects. We derive closed-form expressions for all six nonlinear thermoelectric conductivities and establish a comprehensive symmetry classification under parity and time-reversal operations. Crucially, we show that all extrinsic scattering contributions have a band geometric origin and are universally proportional to the Berry curvature.  

Our symmetry analysis reveals that asymmetric-scattering driven nonlinear thermoelectric responses are naturally enhanced in ${\cal C}_3$-symmetric systems, including graphene multilayers. As a concrete case study, we apply our framework to ABA-stacked trilayer graphene and show that extrinsic skew-scattering mechanisms dominate the nonlinear Nernst and Seebeck effects, consistent with the giant signals observed experimentally~\cite{liu2025nonlinear}. Altogether, our results establish the comprehensive theory of extrinsic nonlinear thermoelectricity. By clarifying how band geometry combines with asymmetric impurity scattering to generate significant nonlinear thermoelectric responses, our work outlines a pathway toward high-efficiency, magnetic-field-free thermoelectric functionality. 
  
\section{Extrinsic nonlinear Nernst and Seebeck Effects}

We now develop a semiclassical kinetic framework that systematically incorporates disorder-induced asymmetric scattering. The goal of this section is to derive the general expression for the second-order thermoelectric current under a temperature gradient and identify all distinct contributions arising from intrinsic dynamics, side-jump processes, and skew scattering mechanisms. As we show below, this naturally leads to six distinct nonlinear thermoelectric current channels. These will serve as the basis for the symmetry analysis and material-specific results presented in later sections. 

Within the semiclassical framework, the thermoelectric charge current takes the form~\cite{xiao2007valley, qin2011energy, kamal2021intrinsic, papaj2021enhanced, varshney2025planar},  
\bea \label{eq:j_def} 
    {\bm j} &=& -e \sum_l \dot{\bm r} f_l - \frac{e}{\hbar} \frac{\gradient_{\bm r} T}{T}  \times \sum_{l}  {\bm \Omega}_l \bigg[ (\e_l - \mu) f_l  \nn \\ 
&& + k_B T \log\{ 1 + {\rm exp}(-\beta(\e_l - \mu))\}\bigg]~. 
\eea
Here, $l = (n, \kb)$ labels a Bloch state in band $n$ with momentum $\kb$, $-e~(e>0)$ is the electron charge, ${\bm \Omega}_l$ is the Berry curvature, and $T \equiv T(\bm r)$ is the local temperature. Additionally, $f_l ({\bm r}, {\bm k})$ is the nonequilibrium distribution function, $\e_l$ is the band energy of $l$-th band, $\mu$ is the chemical potential, $\beta = 1/k_B T$, and $\gradient_{\bm r} T$ is the temperature gradient. The first term in Eq.~\eqref{eq:j_def} represents the current from the wavepacket velocity $\dot{\bm r}$. The second term is the magnetization current arising from the finite size of the electron wavepacket \cite{Burgos_Atencia31122024}. The notation  $\sum_{l} = \mathscr{V}\sum_n \int [d\kb]$ denotes both band summation and momentum integration, with $[d\kb] = (d^d\kb)/(2\pi)^d$ in $d$-dimensions and $\mathscr{V}$ being the real-space volume of the system.
For calculating the nonlinear thermoelectric current, we need to estimate the wavepacket velocity $\dot{\bm r}$ and the nonequilibrium distribution function $f_l$.   

Disorder gives rise to an additional contribution to the effective velocity of the electron wavepacket~\cite{papaj2021enhanced, zhou2022fundamental}. We have,
\be
\dot{\bm r}_l = {\bm v}_l + {\bm v}_l^{\rm sj}~,
\ee
where, ${\bm v}_l = \hbar^{-1} \gradient_{\kb} \e_l$ is the band velocity, and ${\bm v}_l^{\rm sj} = - \sum_{l'} w_{ll'}^{S} \delta {\bm r}_{ll'}$ is the side-jump velocity contribution. Here, $w_{ll'}^{S}$ is the symmetric part of the scattering rate, while $\delta {\bm r}_{ll'}$ is the real-space positional shift of the Bloch electron during a scattering event. 
In addition to the velocity contribution, scattering modifies the nonequilibrium distribution function. 

\subsection{Nonequilibrium distribution function}
To calculate the nonequilibrium distribution function under a temperature gradient, we use the semiclassical Boltzmann transport equation~\cite{ashcroft1976ssp}, 
$\frac{\partial f_l}{\partial t} + \dot{\bm r}\cdot \gradient_r f_l = I_{\rm el}(f_l)$. Here, $I_{\rm el}(f_l)$ is the elastic collision integral describing scattering of Bloch electrons from static impurities~\cite{papaj2021enhanced}, 
\be\label{eq:col_int}
I_{\rm el}(f_l) = -\sum_{l'} \big[w_{ll'} f_l({\bm r}) - w_{l'l} f_{l'} ({\bm r} + \delta{\bm r}_{ll'})\big]~.
\ee
The scattering rate $w_{ll'}$ denotes the transition of an electronic Bloch state from $l$ to $l'$, which is governed by the Fermi's Golden rule~\cite{ashcroft1976ssp},  
\be\label{eq:sr_def}
w_{ll'} = \frac{2\pi}{\hbar}\langle \vert \bra{u_l} V_{\rm imp}\ket{\Psi_{l'}}\vert^2\rangle_{\rm dis} \delta({\e_l -\e_{l'}})~.
\ee
Here, $\ket{u_l}$ is the cell-periodic part of the Bloch state of the disorder-free unperturbed Hamiltonian ${\m H}_0 (\kb)$,  $V_{\rm imp}$ is the impurity potential, and $\ket{\Psi_{l'}}$ is the eigenstate of the full Hamiltonian ${\m H} (\kb) = {\m H}_0 (\kb) + V_{\rm imp}$. In Eq.~\eqref{eq:sr_def}, disorder averaging is denoted by $\langle \cdots\rangle_{\rm dis}$. $\ket{\Psi_{l'}}$ can be expanded iteratively using the Lipmann-Schwinger  equation~\cite{sakurai2017QM} 
\be\label{eq:lip_sch}
\ket{\Psi_l} = \ket{u_l} + (\e_l - {\m H}_0 + i\eta)^{-1} V_{\rm imp} \ket{\Psi_l}~,
\ee 
with $\eta \to 0^{+}$ ensuring outgoing boundary conditions.  
Additionally, scattering also induces a real-space positional shift of the Bloch electrons during the scattering from state $\ket{u_l}$ to $\ket{u_{l'}}$, which is given by~\cite{sinitsyn2006coordinate} $\delta {\bm r}_{ll'} = \bra{u_l} i \partial_{\kb} u_{l} \rangle - \bra{u_{l'}} i \partial_{\kb'} u_{l'} \rangle - {\bm D}_{\kb \kb'} {\rm arg}(\bra{u_{l}} u_{l'} \rangle)$. ${\bm D}_{\kb \kb'} = \partial_{\kb} + \partial_{\kb'}$ and $\partial_{\kb} \equiv \gradient_{\kb}$, and `arg' denotes the argument of a complex number. This shift plays a central role in side-jump contributions to charge transport. 

In general, elastic scattering processes are not symmetric ($w_{ll'} \ne w_{l'l}$). It is therefore convenient to decompose the scattering rate into symmetric ($w^{S}_{ll'}$) and antisymmetric ($w^{A}_{ll'}$) parts, 
\be\label{eq:sy_nd_as_sr}
w^S_{ll'} = \frac{w_{ll'} + w_{l'l}}{2} = w^S_{l'l}, \ \ \ {\rm and } \ \ \ w^A_{ll'} = \frac{w_{ll'} - w_{l'l}}{2} = - w^A_{l'l}~.
\ee 
The symmetric part $w^S_{ll'}$ governs the conventional relaxation, which is usually treated using the relaxation-time approximation~\cite{ashcroft1976ssp}. The antisymmetric part $w^A_{ll'}$, which is commonly known as the skew-scattering rate, generates skew-scattering contributions to both longitudinal and transverse transport. See Appendix~\ref{app:assymmetric_SR} for more details, where the skew-scattering rate is systematically decomposed into the third- and fourth-order contributions, $w^{(3), A}_{ll'}$ and $w^{(4), A}_{ll'}$, respectively. The third-order skew-scattering rate $w^{(3), A}_{ll'}$ arises from the third-order Born approximation of the impurity potential, whereas the fourth-order skew-scattering rate $w^{(4), A}_{ll'}$ originates from the fourth-order Born approximation. 

Using the above decomposition, and a first order expansion of $f_{l'}({\bm r} + \delta{\bm r}_{ll'}) \approx f_{l'}({\bm r}) + \delta{\bm r}_{ll'}\cdot\gradient_{\bm r} f_{l'}({\bm r})$, the elastic collision integral separates naturally into three channels, intrinsic, side-jump, and skew-scattering, each with distinct physical origin. We have, $$I_{\rm el}(f_l) = I^{\rm in}_{\rm el}(f_l) + I^{\rm sj}_{\rm el}(f_l) + I^{\rm sk}_{\rm el}(f_l)~,$$ and  
these distinct components have the form~\cite{du2019disorder}, 
\begin{subequations}
\begin{align}
    I^{\rm in}_{\rm el}(f_l) &= -\sum_{l'} w^{S}_{ll'}(f_l - f_{l'})~, \\ 
    I^{\rm sj}_{\rm el}(f_l) &= \sum_{l'} w^S_{ll'}\delta{\bm r}_{ll'}\cdot \gradient_{\bm r}f_{l'}~, \\ 
    I^{\rm sk}_{\rm el}(f_l) &= \sum_{l'} w^{A}_{ll'}(f_l + f_{l'})~.
\end{align}
\end{subequations}
Here, $I^{\rm in}_{\rm el}$ captures intrinsic relaxation by symmetric scattering, $I^{\rm sj}_{\rm el}$ represents the side-jump contributions, and $I^{\rm sk}_{\rm el}$ encodes skew-scattering collision integral. Under weak disorder, 
$w^A_{ll'} \ll w^S_{ll'}$, so subleading mixed terms like $w^A_{ll'} \delta {\bm r}_{ll'}$ in the side-jump channel can be safely neglected. This decomposition provides a systematic way to study intrinsic, side-jump, and skew-scattering contributions to thermoelectric transport.

Building on the above discussion, we now calculate the solution to the Boltzmann equation in the perturbative limit. We consider a constant, time-independent temperature gradient and focus on the steady state distribution function with $\partial_t f_l = 0$. Analogous to the elastic collision integral, we decompose the nonequilibrium distribution function into intrinsic, side-jump, and skew-scattering parts, $f_l = f^{\rm in}_l + f^{\rm sj}_l + f^{\rm sk}_l$. 
Here, $f_l^{\rm sj}$ and $f_l^{\rm sk}$ represent the disorder-induced corrections that depend explicitly on the applied temperature gradient $\gradient_{\bm r} T$. In contrast, $f^{\rm in}_l$ consists of an local equilibrium contribution, $f^0_l = [1 + {\rm exp}(\beta(\e_l - \mu))]^{-1}$ and a temperature gradient-induced contribution. Using the above decomposition and $\gradient_{\bm r} f_l \equiv \gradient_{\bm r} T~ \partial f_l/\partial T$, into the Boltzmann equation, we obtain the following set of self-consistent relations, 
\begin{subequations}\label{eq:sbe}
\begin{align}
\dot{\bm r}\cdot \gradient_{\bm r} T \pdv{f^{\rm in}_l}{T} &= I^{\rm in}_{el} (f^{\rm in}_{l})~,\label{eq:sbe_int} \\ 
\dot{\bm r}\cdot \gradient_{\bm r} T \pdv{f^{\rm sj}_l}{T} &= I^{\rm in}_{el} (f^{\rm sj}_{l}) + I^{\rm sj}_{el} (f^{\rm in}_{l}) ~,\label{eq:sbe_sj} \\
\dot{\bm r}\cdot \gradient_{\bm r} T \pdv{f^{\rm sk}_l}{T} &= I^{\rm in}_{el} (f^{\rm sk}_{l}) + I^{\rm sk}_{el} (f^{\rm in}_{l}) ~. \label{eq:sbe_sk}
\end{align}
\end{subequations}
These equations form the basis for the steady state distribution function under a temperature gradient with disorder. Following Ref.~\cite{du2019disorder}, we neglect the subleading mixed side-jump–skew contributions. 

We solve these coupled equations explicitly in  Appendix~\ref{app:sbe_sol}, to derive the first- and second-order corrections to the nonequilibrium distribution functions. Following that, we use them to calculate the extrinsic contributions to nonlinear thermoelectric currents. As we demonstrate below, the second-order thermoelectric current naturally decomposes into six distinct contributions. Four of these six contributions originate from asymmetric scattering mechanisms. Crucially, we show in Sec. IV that for short-range disorder, these extrinsic contributions are proportional to the Berry curvature. This reveals their band geometric dependence even though they arise from disorder.

\subsection{Nonlinear Nernst and Seebeck currents}
\begin{table*}[t]
\caption{ Summary of the six second-order nonlinear thermoelectric conductivity contributions.
Each row lists the explicit conductivity tensor, its microscopic ingredients, its dependence on Berry curvature ($\propto \bm \Omega$) and antisymmetric scattering rate ($\propto w^A$), and symmetry constraints under parity ($\mathcal P$) and time-reversal ($\mathcal T$) symmetries. Below, we have defined $\tilde{\e}_l = (\e_l - \mu)$ for brevity. Here, $\cancel{\mathcal P}$ and $\cancel{\mathcal T}$ denote broken parity and time-reversal symmetry, respectively. The last column indicates whether the response contributes to the nonlinear Nernst effect (NNE), the nonlinear Seebeck effect (NSE), or both.}
\label{tab:NL_eq_sym_summary}
\centering
\renewcommand{\arraystretch}{2.5}
\setlength{\tabcolsep}{5pt}
\begin{tabular}{c p{8.5cm} c c c c c c}
\hline\hline
\textbf{Response}
& \textbf{Conductivity tensor $\alpha_{a;bc}$}
& $\propto \boldsymbol{\Omega}$
& $\propto w^A$
& $\cancel{\mathcal P}\cancel{\mathcal T}$
& $\mathcal P,\cancel{\mathcal T}$
& $\mathcal T,\cancel{\mathcal P}$
& NNE / NSE \\
\hline\hline
$\alpha^{\rm ND}$
&
$-e\tau^2 \sum_l v_l^a v_l^b v_l^c \, \partial_T^2 f_l^0$
& \no & \no
& \yes & \no & \no
& NSE \& NNE
\\
$\alpha^{\rm NA}$
&
$\dfrac{e\tau}{2\hbar T}\sum_l \tilde{\e}_l
(\Omega_l^{ab} v_l^c + \Omega_l^{ac} v_l^b)\, \partial_T f_l^0$
& \yes & \no
& \yes & \no & \yes
& NNE
\\
$\alpha^{\rm NSJ}$
&
$-e\tau^2 \sum_l
\!\left(v^{\rm sj,a}_l v_l^b v_l^c
+ v_l^a v^{\rm sj,b}_l v^{\rm sj,c}_l\right)\!
\partial_T^2 f_l^0$
& \yes & \yes
& \yes & \no & \yes
& NNE \& NSE
\\
$\alpha^{\rm NSK}$
&
$-\dfrac{e\tau^3}{2}\!\sum_{l,l'}\! w^A_{ll'} \partial_T^2 f_l^0
\!\left[4v_l^a v_l^b v_l^c
-2v_{l'}^a v_l^b v_l^c
- v_{l'}^a v_{l'}^b v_l^c
- v_{l'}^a v_{l'}^c v_l^b\right]$
& \yes & \yes
& \yes & \no & \yes
& NNE \& NSE
\\
$\alpha^{\rm NASJ}$
&
$\dfrac{e\tau}{2\hbar T}\sum_l \tilde{\e}_l
(\Omega_l^{ab} v^{\rm sj,c}_l + \Omega_l^{ac} v^{\rm sj,b}_l)\,
\partial_T f_l^0$
& \yes & \yes
& \yes & \no & \no
& NNE
\\
$\alpha^{\rm NASK}$
&
$\dfrac{e\tau^2}{2\hbar T}\!\sum_{l,l'}\! w^A_{ll'} \partial_T f_l^0
\!\left[\tilde{\e}_l(\Omega_l^{ab} v_l^c + \Omega_l^{ac} v_l^b)
- \tilde{\e}_{l'}(\Omega_{l'}^{ab} v_l^c + \Omega_{l'}^{ac} v_l^b)\right]$
& \yes & \yes
& \yes & \no & \no
& NNE  \\[1ex]
\hline\hline
\end{tabular}
\end{table*}

To evaluate the second-order nonlinear thermoelectric current, we expand the distribution function $f_l$ up to quadratic order in Eq.~\eqref{eq:j_def} and retain the terms proportional to $(\nabla T)^2$. This yields
\begin{widetext}
\be
 {\bm j}^{(2)} = - e \sum_l ({\bm v}_l + {\bm v}^{{\rm sj}}_l) \left[f^{\rm in, (2)}_l + f^{\rm sj, (2)}_l + f^{\rm sk, (2)}_l\right] - \frac{e}{\hbar} \frac{\gradient_{\rm r} T}{T} \times \sum_l {\bm \Omega}_l (\e_l - \mu) \left[f^{\rm in, (1)}_l + f^{\rm sj, (1)}_l + f^{\rm sk, (1)}_l\right]~. \nn
\ee 
\end{widetext}
This expression naturally separates into six distinct channels, ${\bm j}^{(2)} = {\bm j}^{\rm ND} + {\bm j}^{\rm NA} + {\bm j}^{\rm NSJ} + {\bm j}^{\rm NSK} + {\bm j}^{\rm NASJ} + {\bm j}^{\rm NASK} $. Here, ${\bm j}^{\rm ND}$ is the nonlinear Drude (ND) current from the band velocity; ${\bm j}^{\rm NA}$ is the nonlinear anomalous (NA) current associated with the Berry curvature; ${\bm j}^{\rm NSJ}$ is the nonlinear side-jump (NSJ) current stemming from the side-jump velocity; and ${\bm j}^{\rm NSK}$ is the nonlinear skew-scattering (NSK) current driven by the antisymmetric part of the scattering rate. Following Ref.~\cite{ma2023anomalous}, we label the remaining two hybrid terms (${\bm j}^{\rm NASJ}$ and ${\bm j}^{\rm NASK}$) as the nonlinear anomalous side-jump (NASJ) and nonlinear anomalous skew-scattering (NASK) currents, since these involve both Berry curvature and disorder effects. The NASJ current combines Berry curvature with side-jump velocity, while the NASK current arises from Berry curvature and the skew-scattering rate. The explicit forms of these contributions are, 
\begin{subequations}\label{eq:NLC_parts}
    \begin{align}
        {\bm j}^{\rm ND} & = -e \sum_l {\bm v}_l f^{\rm in, (2)}_l~, \\ 
        {\bm j}^{\rm NA} & = -\frac{e}{\hbar} \frac{\gradient_{\bm r} T}{T} \times \sum_l {\bm \Omega}_l (\e_l - \mu) f^{\rm in, (1)}_l~, \\
        {\bm j}^{\rm NSJ} &= -e \sum_l \left[{\bm v}_l f^{\rm sj, (2)}_l + {\bm v}^{\rm sj}_l f^{\rm in, (2)}_l\right]~, \\ 
        {\bm j}^{\rm NSK} &= - e \sum_l {\bm v}_l f^{\rm sk, (2)}~, \\
        {\bm j}^{\rm NASJ} &= - \frac{e}{\hbar} \frac{\gradient_{\bm r} T}{T} \times \sum_l {\bm \Omega}_l (\e_l - \mu)f^{\rm sj, (1)}_l~, \\
        {\bm j}^{\rm NASK} &= - \frac{e}{\hbar} \frac{\gradient_{\bm r} T}{T} \times \sum_l {\bm \Omega}_l (\e_l - \mu) f^{\rm sk, (1)}_l~.
    \end{align}
\end{subequations}
By substituting the full expression of $f^{(1)}_l$ and $f^{(2)}_l$ from Appendix~\ref{app:sbe_sol} into the above equations, one can write the second-order thermoelectric current in compact form as $j^{(2)}_a = \alpha_{a;bc} \nabla_b T \nabla_c T$, where $\alpha_{a;bc}$ is the third-rank nonlinear thermoelectric conductivity tensor, and $a$, $b$, and $c$ denote Cartesian indices. Because the temperature gradient components commute $i.e., ~ \nabla_b T \nabla_c T = \nabla_c T \nabla_b T$, therefore  conductivity tensor $\alpha_{a;bc}$ must be symmetric in field indices ($b,~c$). This symmetry can be easily enforced by replacing $\alpha_{a;bc}$ with $(\alpha_{a;bc} + \alpha_{a;cb})/2 $. Following Eq.~\eqref{eq:NLC_parts}, $\alpha_{a;bc}$ naturally decomposes into six distinct contributions. We present the resulting simplified expression of these contributions in Table~\ref{tab:NL_eq_sym_summary}. 

Together, these describe all contributions to second-order nonlinear Nernst, Seebeck, and mixed thermoelectric responses. The Drude and nonlinear anomalous conductivities, $\alpha^{\rm ND}_{a;bc}$ and $\alpha^{\rm NA}_{a;bc}$ reproduce results consistent with earlier studies~\cite{zeng2019nonlinear, yu2019topological,Wu2021nonlinear,varshney2025intrinsic}. The additional asymmetric scattering contributions $\alpha^{\rm NSJ}_{a;bc}$, $\alpha^{\rm NSK}_{a;bc}$, $\alpha^{\rm NASJ}_{a;bc}$, and $\alpha^{\rm NASK}_{a;bc}$ have not been explored thoroughly earlier and these are the main findings of this paper. Since all these conductivities depend on the derivatives of the Fermi–Dirac distribution function ($f^0_l$), they are genuine Fermi-surface effects, which manifest in metallic systems with a finite density of states at the Fermi energy. 

Importantly, while the $\alpha^{\rm ND}_{a;bc}$, $\alpha^{\rm NSJ}_{a;bc}$, and $\alpha^{\rm NSK}_{a;bc}$ channels contribute to both NNE and NSE, the anomalous channels $\alpha^{\rm NA}_{a;bc}$, $\alpha^{\rm NASJ}_{a;bc}$, and $\alpha^{\rm NASK}_{a;bc}$ contribute exclusively to NNE,  since they originate from the cross product of the Berry-curvature and temperature gradient ($\nabla_{\bm r}T \times {\bm \Omega}_{\kb}$). As a result, the NSE is governed entirely by the ND, NSJ, and NSK contributions, while the NNE uniquely probes additional disorder effects driven by the Berry curvature. 

To understand the temperature dependence of these thermoelectric responses better, it is useful to express the temperature derivatives of the equilibrium Fermi-Dirac distribution function as
\begin{subequations}\label{eq:T_der_ff}
    \begin{align}
        \pdv{f^0_l}{T} &= -\frac{ \tilde{\e}_l}{T} \pdv{f^0_l}{\e_l}~, \\ 
        \pdv[2]{f^0_l}{T} &= \frac{ \tilde{\e}_l}{T^2} \left[2\pdv{f^0_l}{\e_l} + \tilde{\e}_l \pdv[2]{f^0_l}{\e_l}   \right]~.  
    \end{align}
\end{subequations}
These relations indicate how thermal broadening near the Fermi surface governs the temperature dependence of all six conductivity channels. 
%

\section{Symmetry analysis of the responses}

\begin{table}[t!]
\caption{Transformation of momentum dependent physical quantities under parity ($\mathcal P$) and time-reversal ($\mathcal T$) symmetry. Here, $\cancel{\mathcal P}$ and $\cancel{\mathcal T}$ denote broken parity and time-reversal symmetry, respectively.}
\centering
    \renewcommand{\arraystretch}{1.5} 
    \setlength{\tabcolsep}{10pt}
    \begin{tabular}{ c c c }
    \hline\hline
   \textbf{Quantities}&  $ {\mathcal P},~\cancel{\mathcal T}$& ${\mathcal T},~\cancel{\mathcal P}$ \\ 
    \hline\hline
    ${\bm k}$ & $-\kb$ & $-\kb$ \\ 
    $\e(\kb)$ & $\e(-\kb)$ & $\e(-\kb)$ \\ 
    ${\bm v}(\kb)$ & $-{\bm v}(-\kb)$ & $-{\bm v}(-\kb)$ \\ 
    $\delta{\bm r}(\kb,\kb')$ & $-\delta{\bm r}(-\kb,-\kb')$ & $\delta{\bm r}(-\kb, -\kb')$ \\ 
    ${\bm v}^{\rm sj}(\kb)$ & $-{\bm v}^{\rm sj}(-\kb)$ & ${\bm v}^{\rm sj}(-\kb)$ \\ 
    ${\bm \Omega}(\kb)$ & ${\bm \Omega}(-\kb)$ & $- {\bm \Omega}(-\kb)$ \\
    $w^S(\kb, \kb') $ & $w^S(-\kb, -\kb') $ & $w^S(-\kb, -\kb') $ \\ 
    $ w^A(\kb, \kb')$ & $ w^A(-\kb, -\kb')$ & $ -w^A(-\kb, -\kb')$ \\ 
    \hline\hline
    \end{tabular}
    \label{tab:PQ_sym}
\end{table}

Having derived the six distinct nonlinear thermoelectric conductivity channels, we now analyze the constraints imposed on these by fundamental symmetries. Since each contribution is constructed from momentum-dependent quantities such as velocities, Berry curvature, and asymmetric scattering rates, symmetry constraints can be determined by examining how these constituents transform under parity ($\cal P$) and time-reversal ($\cal T$) operations. We present the transformation of the momentum-dependent constituents under these symmetry operations in Table~\ref{tab:PQ_sym}. The table highlights whether each component is odd (changes sign) or even (remains invariant) under the corresponding symmetry transformation. Since all responses involve Brillouin-zone integrals, any integrand that is odd under $\bm k \to -\bm k$ vanishes identically. Thus, Table~\ref{tab:PQ_sym} provides a symmetry-based criterion for determining which response channels survive.

Under inversion $(\mathcal P)$ symmetry, all nonlinear thermoelectric conductivities are odd under $\bm k \to -\bm k$ and therefore vanish identically in centrosymmetric systems. In inversion broken but time-reversal-preserving systems ($\mathcal T$), we find that $\alpha^{\rm ND}$, $\alpha^{\rm NASJ}$, and $\alpha^{\rm NASK}$ are $\bf k$-odd and thus forbidden, whereas $\alpha^{\rm NA}$, $\alpha^{\rm NSJ}$, and $\alpha^{\rm NSK}$ are $\bf k$-even and yield finite responses. 

Importantly, the nonlinear conductivities  $\alpha^{\rm NASJ}$ and $\alpha^{\rm NASK}$, are symmetry-allowed only when both inversion and time-reversal symmetries are individually broken. These responses can therefore be used as a symmetry-based diagnostic tool to identify systems hosting non-centrosymmetric magnetic order.
Representative examples include altermagnets (e.g., RuO$_2$, MnTe), non-centrosymmetric ferromagnets (e.g., MnSi, FeGe), non-centrosymmetric antiferromagnets (e.g., Mn$_3$Sn, Mn$_3$Ge), and layered van der Waals magnets lacking inversion symmetry.

\section{Band-geometric origin of extrinsic contributions}\label{sec:disorder_info}
The side-jump velocity and antisymmetric scattering (or skew-scattering) rate are the central microscopic ingredients underlying extrinsic nonlinear responses. Both quantities arise from overlaps between Bloch states at different crystal momenta, with the intraband Berry connection appearing explicitly in the side-jump velocity through the positional shift. This naturally raises an important question: do extrinsic responses also encode geometric information of the Bloch bands? To address this question, we need a model for the impurity potential.

\subsection{Impurity potential modeling}
To evaluate extrinsic responses associated with the side-jump velocity and skew-scattering mechanism, we consider randomly distributed, short-range static scatterers following Refs.~\cite{du2019disorder, papaj2021enhanced, ma2023anomalous, guo2024extrinsic, ma2025quantum}. For concreteness, we  model them as a Dirac-delta potential, 
\be
V_{\rm imp}({\bm r}) = \sum_j V_j \delta({\bm r} - {\bm R}_j)~.
\ee
Here, $V_j$ denotes the impurity strength at site ${\bm R}_j$, and the summation runs over all impurity locations. For this impurity potential, the matrix element of impurity potential in Bloch basis is given by $ V_{ll'} \equiv V^0_{\kb, \kb'} \langle u_l \vert u_{l'} \rangle$, where $ V^0_{\kb, \kb'} = \sum_j V_j e^{i(\kb' - \kb)\cdot {\bm R}_j}$ denotes the Fourier transformation of the impurity potential. This leads to the following relations 
\bea 
\langle V_{ll'
} V_{l'l} \rangle_{\rm dis} &=& \langle V^0_{\kb \kb'} V^0_{\kb'\kb}  \rangle_{\rm dis} \langle u_l \vert u_{l'} \rangle \langle u_{l'} \vert u_{l} \rangle ~, \\
\langle V_{ll''
} V_{l''l'} V_{l'l} \rangle_{\rm dis} &=& \langle  V^0_{\kb \kb''}  V^0_{\kb'' \kb'} V^0_{\kb' \kb} \rangle_{\rm dis} \times \nn \\
&& \langle u_l \vert u_{l''} \rangle \langle u_{l''} \vert u_{l'} \rangle \langle u_{l'} \vert u_l \rangle~, \nn \\
\langle V_{ll''
} V_{l''l'} V_{l'l'''} V_{l'''l} \rangle_{\rm dis} &=& \langle  V^0_{\kb \kb''}  V^0_{\kb'' \kb'}  V^0_{\kb' \kb'''}  V^0_{\kb''' \kb} \rangle_{\rm dis}  \times \nn \\ 
&& \langle u_l \vert u_{l''} \rangle \langle u_{l''} \vert u_{l'} \rangle \langle u_{l'} \vert u_{l'''} \rangle \langle u_{l'''} \vert u_l \rangle~. \nn 
\eea 
Furthermore, we consider the disorder averages to be $\langle V^0_{\kb \kb'} V^0_{\kb'\kb}  \rangle_{\rm dis} = n_i V_0^2$, $\langle  V^0_{\kb \kb''}  V^0_{\kb'' \kb'} V^0_{\kb' \kb} \rangle_{\rm dis} = n_i V_1^3$ and $\langle  V^0_{\kb \kb''}  V^0_{\kb'' \kb'}  V^0_{\kb' \kb'''}  V^0_{\kb''' \kb} \rangle_{\rm dis} = n_i^2 V_0^4$. Here, $n_i$ is the impurity concentration and $V_0$ and $V_1$ are the effective moments of the impurity potential, which are also known as the Gaussian and non-Gaussian impurity potential components, respectively.

To use this impurity potential in our numerical results for ABA trilayer graphene, we need realistic numerical values for these parameters. Following Ref.~\cite{ma2023anomalous}, we consider $n_i \approx 10^{10}~{\rm cm^{-2}}$ and $V_0 = 6.2 \times 10^{-13}~{\rm eV~cm^2}$. For these parameters, the symmetric scattering time is $\tau \approx 10^{-13}~{\rm s} $ or $\tau = 0.1$ ps. Similar disorder parameters are also considered in Refs.~\cite{papaj2021enhanced, xiao2021lorentz}. 
To have comparable magnitudes of the nonlinear skew-scattering conductivities originating from the third- and fourth-order skew-scattering rates, we set $V_1 = 0.5 V_0$.

\subsection{Berry curvature dependence of extrinsic contributions for weak disorder}

We analyze the side-jump velocity and skew-scattering rate in the weak-disorder limit. A detailed analytical treatment reveals a direct and explicit connection between these extrinsic quantities and the Berry curvature. Following a systematic simplification, presented in detail in Appendix~\ref{app:simplification}, we obtain compact expressions for the side-jump velocity and antisymmetric scattering rate in terms of the Berry curvature. Our calculation yields, 
\begin{subequations}\label{eq:ext_quantities_in_bc_form}
    \begin{align}
&{\bm v}^{\rm sj}_{n} = \frac{2\pi}{\hbar} n_i V_0^2 \sum_{\kb'} [ (\kb - \kb') \times {\bm \Omega}_n(\kb) ] \delta(\e^\kb_n -\e^{\kb'}_n )~,\label{eq:sj_vel_in_bc_form}\\
&w^{(3), A}_{n, \kb \kb'}  = -\frac{2\pi^2}{\hbar} n_i V_1^3 \sum_{\kb''} \delta(\e^\kb_n - \e^{\kb'}_n)\delta(\e^\kb_n - \e^{\kb''}_n)  \\ 
& \quad \quad \quad\times [ (\kb \times \kb'')  + (\kb'' \times \kb') +  (\kb' \times \kb) ] \cdot {\bm \Omega}_n~,\nn \\
& w^{(4), A}_{n, \kb \kb'} =   -\frac{2 \pi^2}{\hbar} n_i^2 V_0^4 \sum_{\kb''} \delta(\e_n^\kb - \e_n^{\kb'}) \delta(\e_n^{\kb'} - \e_n^{\kb''})~ \\ 
& \quad \quad \quad \times  \left[ (\kb \times \kb'') + (\kb'' \times \kb') + (\kb' \times \kb) \right] \cdot \tilde{\bm \Omega}_{n}~. \nn
    \end{align}
\end{subequations} 
Here, $\tilde{\bm \Omega}_n = \sum_{n' \ne n} {\bm \Omega}_{nn'}/(\e^\kb_n - \e^\kb_{n'})$ denotes the energy normalized Berry curvature. These general closed-form expressions highlight that for short-range disorder, the leading extrinsic contributions arising from both the side-jump velocity and the skew-scattering rate are fundamentally governed by the Berry curvature. Moreover, for systems with an even energy dispersion $\e(\kb) = \e(-\kb)$, the above expressions admit further simplification in terms of band-resolved density of states. The resulting simplified forms are presented in Sec.~\ref{app:even_E_form_of_ext_quant}.

These simplified expressions also significantly simplify numerical calculations of extrinsic contributions. By expressing these quantities in terms of the Berry curvature, we avoid the explicit evaluation of interband matrix elements and multiband overlap factors, which are numerically demanding. Additionally, this reformulation enables the efficient and stable computation of nonlinear side-jump and skew-scattering contributions using on-shell energy constraints, which are captured by Dirac delta functions. We now use the developed framework to calculate extrinsic responses in ABA trilayer graphene.

\begin{figure*}[t!]
    \centering
    \includegraphics[width= 0.9\linewidth]{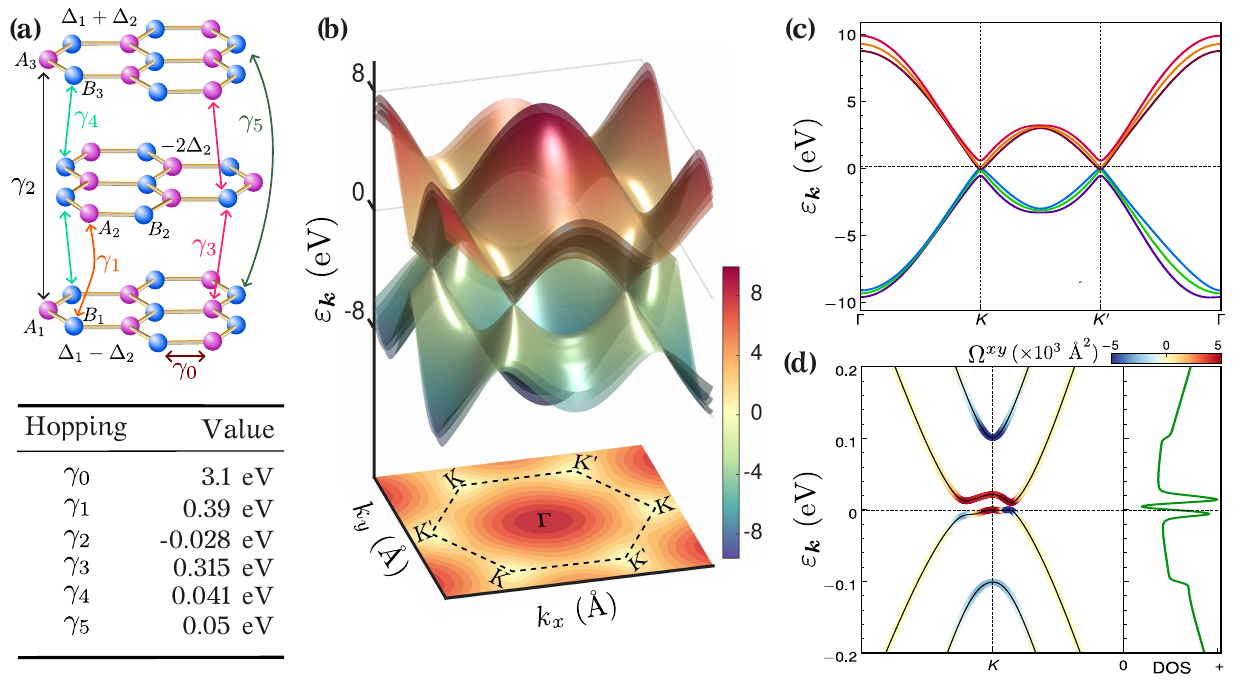}
    \caption{ \textbf{Lattice structure, electronic properties and  Berry curvature of ABA-staked trilayer graphene (TLG).} 
    (a) Crystal structure of ABA TLG with intra- and interlayer hopping parameters listed in the bottom panel. 
    (b) Full band structure from $-10$ to $10$ eV, with the first conduction band projected onto the $k_x$–$k_y$ plane. The high-symmetry points ($\Gamma$, $K$, $K'$) and the first Brillouin zone are marked in black. 
    (c) Band dispersion along high-symmetry lines, highlighting low-energy crossings near the Dirac points ($K$, $K'$). 
    (d) Zoomed view near the $K$ point with Berry curvature projected onto the bands, showing pronounced hotspots near the charge-neutrality point. The right panel shows the corresponding density of states (DOS) as a function of energy. 
    In all plots, we use the onsite potential ($\delta$) and the layer potential due to the applied vertical electric field as ($\Delta_1, ~ \Delta_2$) as $[\delta, \Delta_1, \Delta_2] = [0.046,\ 0.1,\ 0]$ eV.
    \label{fig2}}
\end{figure*}

\section{Large response in ABA stacked trilayer graphene}
\label{Sec_V}
As discussed in the symmetry analysis section, TRS-preserving systems are promising candidates for observing the NNE and NSE, since they permit asymmetric scattering-driven responses. Two-dimensional systems in dual-gated geometry, such as bilayer and trilayer graphene and their moir\'e counterparts, with tunable carrier density and vertical displacement field, are good candidates for exploring these responses. 
We focus on Bernal-stacked (ABA) trilayer graphene (TLG) to examine the nonlinear thermoelectric conductivities. Our choice of TLG is motivated by the recent experimental observation of a giant NNE in ABA stacked TLG~\cite{liu2025nonlinear}. We first discuss the electronic properties of ABA-TLG, followed by NNE in ABA-TLG. 

\begin{figure}[t!]
\centering
\includegraphics[width=\linewidth]{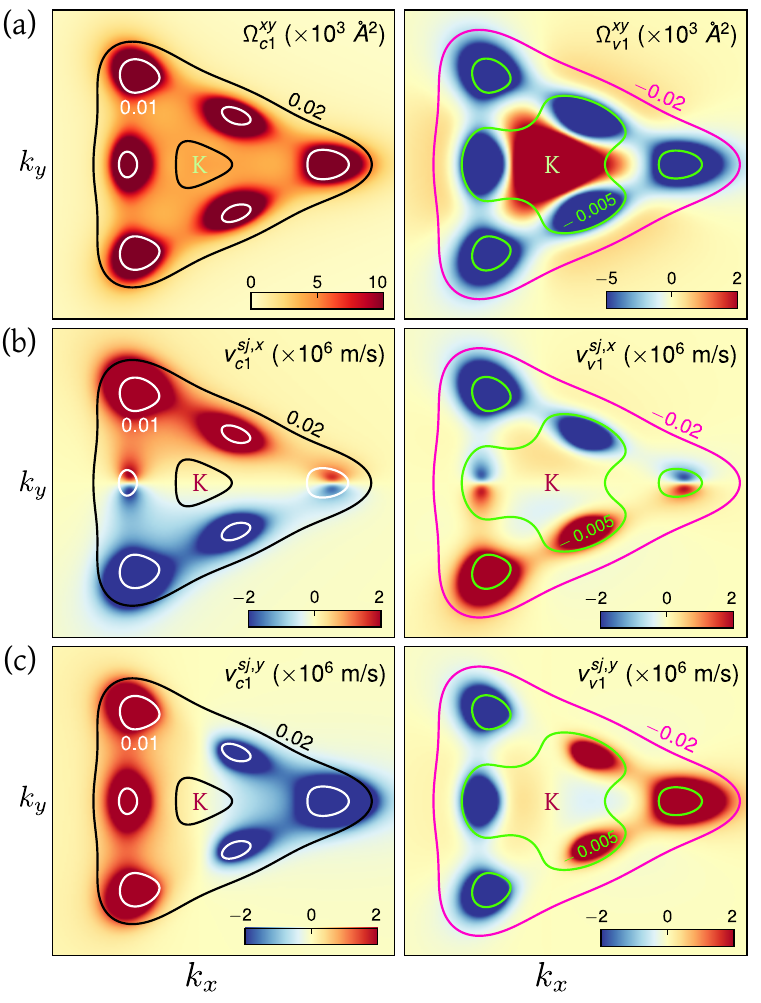}
    \caption{\textbf{Momentum-space distribution of the Berry curvature and side-jump velocity for ABA-TLG.} a) Berry curvature around the $K$ valley for the first conduction (c1) and valence (v1) bands, shown in the left and right panels, respectively. (b-c) The same plotting scheme is used for the $x$ and $y$ components of the side-jump velocity around the $K$ valley. Energy contours are drawn at $0.01$ and $0.02$~eV in the conduction band, and at $-0.005$ and $-0.02$~eV in the valence band. In all plots, the Hamiltonian parameters are identical to those in Fig.~\ref{fig2}. }
    \label{fig:bcq}
\end{figure}

\subsection{Electronic properties and Berry curvature}
Figure~\ref{fig2}(a) illustrates the lattice structure of ABA-TLG, with the intra- and interlayer hopping parameters used to construct the tight-binding Hamiltonian. Purple spheres denote $A_i$ sites of the $A$ sublattice in layer $i$, while blue spheres denote $B_i$ sites of the $B$ sublattice, with the layer index $i= 1,2,3$. The tight-binding parameters $\gamma_0,\cdots,\gamma_5$ capture the different hopping strengths, as depicted in Fig.~\ref{fig2}(a). We have, 
\begin{align}
 A_i \leftrightarrow B_i : \gamma_0~,~ &~ B_{1,3} \leftrightarrow A_2:\gamma_1~,& A_1 \leftrightarrow A_3: \frac{1}{2} \gamma_2~,  
 \nn \\ 
 \quad  A_{1,3} \leftrightarrow B_2: \gamma_3~,~ &~ 
 \begin{array}{l}
 A_{1,3} \leftrightarrow A_2 \\
 B_{1,3} \leftrightarrow B_2 
 \end{array}
 : - \gamma_4~, & B_1 \leftrightarrow B_3: \frac{1}{2} \gamma_5~. \nn
 \end{align}
We list the numerical values of these hoppings in the bottom panel of Fig.~\ref{fig2}(a).

The tight-binding Hamiltonian of ABA-TLG, written in the basis $\{A_1, B_1, A_2, B_2, A_3, B_3\}$, is specified by~\cite{partoens2006from, serbyn2013new} ${\mathcal H}(\kb) = {\mathcal H}_0 + {\mathcal H}_{\Delta_1} + {\mathcal H}_{\Delta_2}$, with
\begin{subequations}\label{eq:Ham}
\begin{align}\label{eq:H_0}
{\mathcal H}_0 &=
\begin{pmatrix}
0 & \gamma_0 t_\kb^* & \gamma_4 t_\kb^* & \gamma_3 t_\kb & \gamma_2/2 & 0 \\[1ex]
\gamma_0 t_\kb & \delta & \gamma_1 & \gamma_4 t_\kb^* & 0 & \gamma_5/2 \\[1ex]
\gamma_4 t_\kb & \gamma_1 & \delta & \gamma_0 t_\kb^* & \gamma_4 t_\kb & \gamma_1 \\[1ex]
\gamma_3 t_\kb^* & \gamma_4 t_\kb & \gamma_0 t_\kb & 0 & \gamma_3 t_\kb^* & \gamma_4 t_\kb \\[1ex]
\gamma_2/2 & 0 & \gamma_4 t_\kb^* & \gamma_3 t_\kb & 0 & \gamma_0 t_\kb^* \\[1ex]
0 & \gamma_5/2 & \gamma_1 & \gamma_4 t_\kb^* & \gamma_0 t_\kb & \delta
\end{pmatrix}~, \\[1ex]
{\mathcal H}_{\Delta_1} &= \text{diag}(\Delta_1, \Delta_1, 0, 0, - \Delta_1, - \Delta_1)~,  \\[1ex] 
{\mathcal H}_{\Delta_2} &= \text{diag}(\Delta_2, \Delta_2, - 2\Delta_2, - 2\Delta_2, \Delta_2, \Delta_2)~.
\end{align}
\end{subequations}
Here, $t_{\bm k} = \sum_{j = 1}^{3} e^{i\kb \cdot a_j} = -1 - 2e^{\sqrt{3}ik_y a/2} \cos(k_x a/2)$ accounts for nearest-neighbor hopping within a monolayer honeycomb lattice. The vectors $\boldsymbol{a}_1 = a(0,~1/\sqrt{3})$, $\bm{a}_2 = a(-1/2,~ -1/2\sqrt{3})$, and $\bm{a}_3 = a(1/2,~ -1/2\sqrt{3})$ connect $A_1$ sites to their nearest neighbors, with $a = 2.46~{\rm \AA}$ being the lattice constant. In addition, $\delta$ is the onsite potential, $2\Delta_1$ is the potential difference between the top and the bottom layers induced by an applied displacement field, and $\Delta_2$ captures the deviation of the middle-layer potential from the average potential of the outer layers. Following Ref.~[\onlinecite{serbyn2013new}], we take $\delta = 0.046$ eV and set $\Delta_2 = 0$, as the latter is typically much smaller than $\Delta_1$.

Diagonalizing ${\mathcal H}(\kb)$ yields the band dispersion of ABA-TLG. Figure~\ref{fig2}(b) shows the full spectrum over an energy window from $-10$~eV to $10$~eV for $\Delta_1 = 0.1$~eV, together with the projection of the first conduction band onto the $k_x$–$k_y$ plane. Figure~\ref{fig2}(c) presents the band structure along high-symmetry lines, highlighting the low-energy regions near the Dirac points ($K$ and $K'$) most relevant for electronic transport. A magnified view around the $K$ point is shown in Fig.~\ref{fig2}(d), along with the distribution of the Berry curvature. The Dirac points act as prominent Berry curvature hotspots, and the Berry curvature decreases rapidly as we move away from $K$. The right panel of Fig.~\ref{fig2}(d) shows the density of states (DOS) as a function of energy $\e_\kb$, in units of $(\text{eV}^{-1} \text{\AA}^{-2})$, where the pronounced peaks corresponds to the van Hove singularities (VHS) near the charge neutrality point (CNP). 

We present the momentum-space distribution of the Berry curvature around the $K$ valley for the first conduction (c1) and first valence (v1) bands, which is displayed in the left and right panels of Fig.~\ref{fig:bcq}(a), respectively. Likewise, fig.~\ref{fig:bcq}(b-c) shows the momentum-space distribution of the $x$ and $y$ components of the side-jump velocity around the $K$ valley for the first conduction and valence bands. As evident from Eq.~\eqref{eq:sj_vel_in_bc_form}, the side-jump velocity inherits its momentum-space structure from the Berry curvature. Therefore, its momentum-space distribution looks qualitatively similar to the Berry curvature distribution. The two Fermi surface contours in all plots, marked by different-colored lines, clearly show the Lifshitz transition in both the conduction and valence bands of ABA-TLG. 


\begin{figure}[t!]
    \centering
    \includegraphics[width= \linewidth]{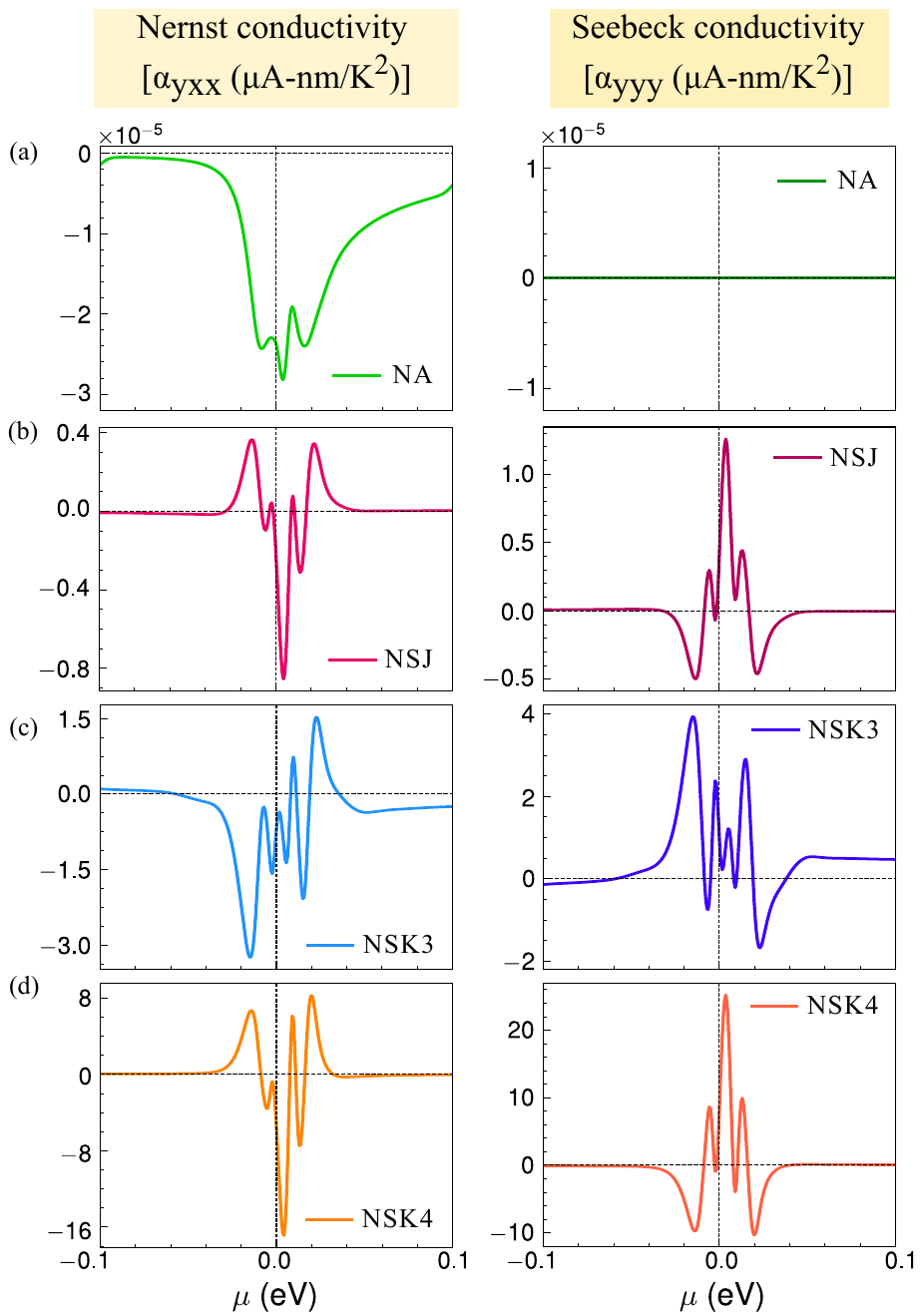}
    \caption{ \textbf{Nonlinear Nernst and Seebeck conductivities in ABA trilayer graphene.} 
    (a–d) Chemical-potential ($\mu$) dependence of the nonlinear Nernst ($\alpha_{yxx}$, left panel) and Seebeck ($\alpha_{yyy}$, right panel) conductivities, expressed in units of ${\rm \mu A~nm/K^2}$. 
    Calculations are performed at $T = 20$ K with symmetric scattering time $\tau = 0.1$ ps, impurity concentration $n_i \approx 10^{10} ~{\rm cm^{-2}}$, $ V_0 = 6.2 \times 10^{-13}~{\rm eV ~cm^2}$ and $V_1 = 0.5 V_0 $. All remaining parameters are the same as in Fig.~\ref{fig2}.}
    \label{fig3}
\end{figure}

\subsection{Large nonlinear Nernst and Seebeck effect}

Using the Berry curvature and electronic properties discussed above in the thermoelectricity conductivity expressions, we numerically calculate the nonlinear Nernst and Seebeck conductivities of ABA-TLG. Since ABA-TLG preserves TRS, Table~\ref{tab:NL_eq_sym_summary} dictates that the ND, NASJ, and NASK contributions vanish identically. Figure~\ref{fig3} shows the remaining NA, NSJ, and NSK components as a function of chemical potential $\mu$. All conductivities are expressed in units of ${\mu}$A-nm/${K^2}$ and the calculations are performed at $T = 20$ K.

These numerical results highlight three salient features. (i) All conductivity components are strongly enhanced near the CNP, reflecting the combined effect of the Berry curvature hotspots and the VHS in the density of states. (ii) The NA contribution originating from the Berry curvature dipole is significantly smaller than the NSJ and NSK terms. 
(iii) The NSK response, which we decomposed into NSK3 and NSK4 as they originate from the third- ($w^{(3),A}_{ll'}$) and fourth-order ($w^{(4),A}_{ll'}$) skew-scattering rates, dominates the nonlinear thermoelectric conductivity.
These findings are in good agreement with the recent experimental observation of Ref.~[\onlinecite{liu2025nonlinear}], which identified skew scattering as the primary source of the giant nonlinear Nernst effect in ABA-TLG.

\subsection{Experimental relevance and comparison} 
To connect our theoretical predictions with experimentally measurable quantities, we convert the computed nonlinear conductivities into nonlinear voltages using~\cite{varshney2025intrinsic}
\be \label{eq:volt_rel}
V^{(2)}_{\perp (||)} =  \frac{\alpha^{(2)}_{\perp (||)} (\nabla T)^2 L_{\perp (||)}}{\sigma^{\rm D}_{||}} ~.
\ee 
Here, $V^{(2)}$ and $\alpha^{(2)}$ denote the nonlinear voltage and nonlinear thermoelectric conductivity, respectively, and the subscripts $\perp$ and $||$ refer to the transverse and longitudinal responses. In Eq.~\eqref{eq:volt_rel}, $L_\perp$ and $L_{||}$ are the device dimensions along the transverse and longitudinal directions, and $\sigma^{\rm D}_{||}$ is the linear Drude conductivity. 

Since all responses are prominent near the CNP, we evaluate the voltages at $\mu = 0.02$ eV, corresponding to the VHS in the DOS. The longitudinal linear Drude conductivity, $\sigma^D_{ab} = - e^2 \tau \int [d\kb] v_a v_b \partial f_0/ \partial \e$ turns out to be $\sigma^{\rm D}_{||} = 0.85~$mA/V for $\mu = 0.02$ eV at $T = 20$ K.
From Fig.~\ref{fig3}, we obtain transverse and longitudinal components of total nonlinear thermoelectric conductivity as $\alpha^{(2)}_{\perp} \approx 9.3~{\rm \mu A~nm/K^2}$ and $\alpha^{(2)}_{||} \approx -11.5~{\rm \mu A ~nm/K^2}$ at $\mu = 0.02$ eV, respectively. Using device dimensions $L_{\perp} = L_{||} \approx 10~\mu$m and an experimentally accessible temperature gradient of $\nabla T = 0.1$ K/$\mu$m~\cite{xu2019large, liu2025nonlinear, hirata2025nonlinear}, Eq.~\eqref{eq:volt_rel} yields nonlinear voltages of the order $\vert V^{(2)}_\perp \vert \approx 1.1~\mu$V and $\vert V^{(2)}_{||}\vert \approx 1.4~\mu$V. These estimated voltage values are well within the experimental observation limit and corroborate the recent experimental findings~\cite{liu2025nonlinear}.

In addition to these nonlinear responses, TLG will also exhibit a conventional linear Seebeck response, primarily driven by the Drude mechanism. Using the expression of the linear Drude thermoelectric conductivity $\alpha^D_{ab} = (e \tau/T) \int [d\kb] (\e - \mu) v_a v_b \partial f_0/ \partial \e$, we obtain $\alpha^D_{||} \approx 13.6$ nA/K at $\mu = 0.02$ eV for temperature 20 K. The corresponding voltage is given by $V^{(1)}_{||} = |\alpha^{(1)}_{||} (\nabla T) L_{||}/ \sigma^D_{||}|$, yields $V^{(1)}_{||} \approx 16~\mu$V. For the parameters we used, the linear Seebeck voltage is nearly 10 times higher than the nonlinear Seebeck voltage. Thus, the nonlinear Seebeck signal has to be carefully extracted either via lock-in frequency measurements or by explicitly subtracting the linear response.  

\begin{figure}[t!]
    \centering
    \includegraphics[width = \linewidth]{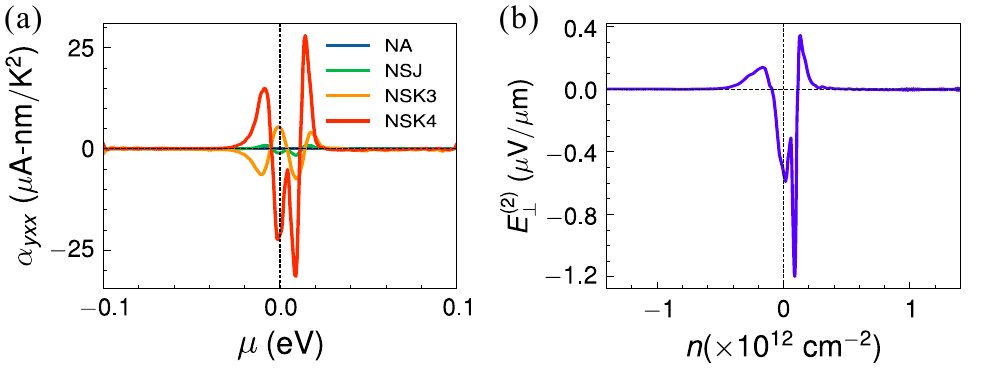}
    \caption{\textbf{Electric field due to the nonlinear Nernst current.} (a) Different components of nonlinear Nernst conductivity are plotted as a function of $\mu$ at a system temperature of 5 K. (b) Variation of the electric field arising due to the nonlinear Nernst current with the electron density ($n$). In computing this, we consider a temperature gradient across the sample of 0.1~K/${\rm \mu m }$ order, while the other system parameters are the same as in the Fig.~\ref{fig3}.}
    \label{fig:exp_comparision}
\end{figure}


To further validate our theoretical results against experimental observations~\cite{liu2025nonlinear}, we compute the electric field corresponding to the nonlinear Nernst current usuing $E^{(2)}_{\perp} = V^{(2)}_{\perp}/L_{\perp} = (\alpha^{(2)}_\perp/\sigma^D_{||}) (\nabla T)^2$, which provides a direct comparison between our theoretical predictions and the experimental observations. In Fig.~\ref{fig:exp_comparision}(a), we present the nonlinear Nernst conductivity as a function of chemical potential computed at a system temperature of 5 K. Using this data, in Fig.~\ref{fig:exp_comparision}(b), we computed $E^{(2)}_{\perp}$ for a temperature gradient 0.1 K/$\rm \mu m$ and convert $\mu$ into electron density $n$ for a direct comparison with the experimental findings. Remarkably, this theoretical estimation closely matches the experimental observation~\cite{liu2025nonlinear}, confirming the reliability and accuracy of our theoretical framework. 

For the impurity potential parameters chosen from Ref.~\cite{ma2023anomalous}, we find that the skew scattering contributions clearly dominate the NNE response. The peak value of the skew scattering contribution is around 20 times larger than the side-jump contribution, and around $10^5$ times that of the anomalous contribution. This raises a natural question: Is there a simple physical explanation to understand this disparity, or does it arise from the specific choice of impurity-scattering potential?

\section{Why Extrinsic Mechanisms Dominate in nearly $C_3$-Symmetric Systems}
To understand the microscopic origin of the large extrinsic NNE and NSE in ABA-TLG, we adopt two approaches. First, we isolate the impurity potential dependence of the nonlinear responses by analyzing their scaled forms. This helps us analyze the dependence of impurity potential and intrinsic material properties separately. Secondly, we examine the crystalline symmetry restrictions on nonlinear thermoelectric responses, which influence the scattering-independent part of the response tensors.  

\subsection{Scaled nonlinear thermoelectric conductivities}
Using the dimensional analysis presented in Appendix~\ref{app:dimension}, we express the nonlinear thermoelectric conductivities in terms of the nonlinear anomalous conductivity. This allows us to quantify the relative strength of the nonlinear side-jump and skew-scattering contributions. We express the ratios, 
\begin{align}
\frac{\alpha^{\rm NSJ}}{\alpha^{\rm NA}} &= N_\tau N_{\rm dis} \frac{\tilde{\alpha}^{\rm NSJ}}{\tilde{\alpha}^{\rm NA}}~, \nn \\ 
\frac{\alpha^{\rm NSK3}}{\alpha^{\rm NA}} &= N^2_\tau {\tilde N}_{\rm dis} \frac{\tilde{\alpha}^{\rm NSK3}}{\tilde{\alpha}^{\rm NA}}~, \\
\frac{\alpha^{\rm NSK4}}{\alpha^{\rm NA}} &= N^2_\tau N^2_{\rm dis} \frac{\tilde{\alpha}^{\rm NSK4}}{\tilde{\alpha}^{\rm NA}}~. \nn
\end{align}
Here, $\tilde{\alpha}$ denotes the dimensionless nonlinear thermoelectric conductivity associated with each channel.

The ratios defined above indicate that the relative magnitudes of the nonlinear conductivities are determined by two key factors. One of these are the dimensionless prefactors associated with the scattering time and impurity potential,
$$ N_{\tau} = \frac{\tau E_s}{\hbar}, \quad N_{\rm dis} = \frac{n_i V_0^2}{E_s^2 a^2}, \quad {\tilde N}_{\rm dis} = \frac{n_i V_1^3}{E_s^3 a^4}~,$$
which encode extrinsic information related to disorder. Here $E_s = 1 $eV is an energy scale, and $a$ is a length scale which can be taken to be the lattice constant. Second, there are the ratios of the dimensionless conductivities $\tilde{\alpha}$, which depend purely on intrinsic electronic properties and symmetries of the system.

In this work, we considered the symmetric scattering time $\tau$ as a constant parameter. Microscopically, it is governed by the symmetric scattering rate~\cite{ashcroft1976ssp}, and it is given by, 
\be
\frac{1}{\tau} = \sum_{l'} w^S_{ll'}(1 - \hat{\kb} \cdot \hat{\kb'}) \quad \implies \tau^{-1} \propto n_i V_0^2~.\nn
\ee 
As a result, the product $N_\tau N_{\rm dis}$ becomes independent of the impurity potential strength and concentration. In contrast, the combination $N_\tau^2 {\tilde N}_{\rm dis} \propto n_i V_1^3 / (n_i^2 V_0^4)$ retains explicit dependence on the impurity potential. 

\begin{figure}[t!]
    \centering
    \includegraphics[width= \linewidth]{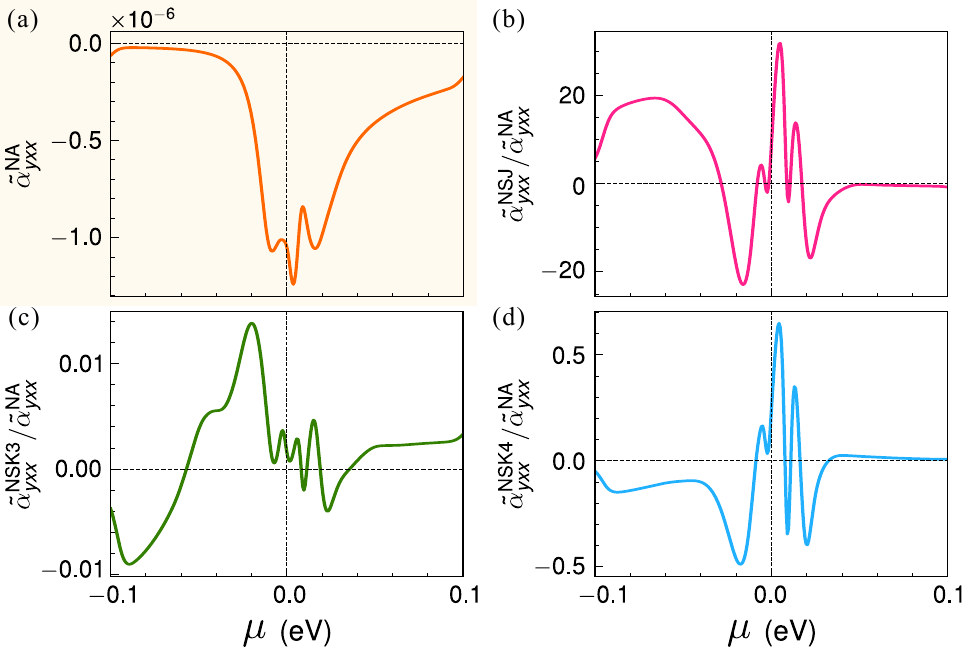}
    \caption{\textbf{Scaled nonlinear Nernst conductivities in ABA-TLG.} (a) Chemical potential dependence of the dimensionless nonlinear anomalous conductivity. (b-d) Presentation of dimensionless nonlinear side-jump  and skew-scattering conductivities in terms of $\tilde{\alpha}^{\rm NA}_{yxx}$.}
    \label{fig:cond_dimensionless}
\end{figure}

To analyze the disorder strength dependence further, we present the `intrinsic' ratios of different contributions in Fig.~\ref{fig:cond_dimensionless}. We find that the intrinsic ratio $\tilde{\alpha}^{\rm NSJ}/\tilde{\alpha}^{\rm NA}$ is substantially larger than the other dimensionless ratios for extrinsic contributions. However, for the chosen disorder parameters, we have 
\be
\begin{aligned}
    N_\tau = 152, \quad N_{\rm dis} = 6.35, \quad {\rm and} \quad {\tilde N}_{\rm dis} = 813.5~, \\ 
    \implies N_\tau N_{\rm dis} = 965.2, \quad {\rm and} \quad  N_{\tau} \tilde{N}_{\rm dis} \approx 1.24 \times 10^5~.  
\end{aligned}\nn
\ee
On combining these prefactors with intrinsic ratios, the nonlinear skew-scattering contribution becomes larger than the nonlinear side-jump contribution. This suggests that the interplay between intrinsic system properties and extrinsic disorder parameters decides the dominating extrinsic channel. In our case, the nonlinear skew-scattering mechanism turns out to be a dominant extrinsic mechanism.

\subsection{Crystalline symmetry restrictions}

Beyond the scaling analysis, we examine the role of crystalline symmetries on the nonlinear thermoelectric response. In addition to TRS, graphene multilayers can also support three-fold rotational symmetry ${\mathcal C}_{3}$, which imposes additional constraints on the tensor elements of $\alpha_{a;bc}$. The tight-binding Hamiltonian of TLG used in Sec.~\ref{Sec_V}, breaks the $C_3$ symmetry by a small amount. As a consequence, we anticipate that TLG will have a small value for those responses that are forbidden by the ${\cal C}_3$ symmetry. 

Let us consider the TLG to be in the $x$–$y$ plane, with an in-plane temperature gradient. In this configuration, the nonlinear thermoelectric conductivity tensor has a total of six independent components: 
two NSE terms $\alpha_{x;xx}$ and $\alpha_{y;yy}$, two NNE terms $ \alpha_{x;yy}, ~\alpha_{y;xx}$, and two mixed contributions 
$ ~\alpha_{x;xy} = \alpha_{x;yx}~, ~\alpha_{y;xy} = \alpha_{y;yx}$. In systems with ${\mathcal C}_3$ symmetry, Newman's theorem forces the following constraints on these
\begin{subequations}\label{eq:C3_sym}
\begin{align}
    &\alpha_{x;yy} = \alpha_{y;xy} = \alpha_{y;yx} = - \alpha_{x;xx}~, \\
    &\alpha_{y;xx} = \alpha_{x;yx} = \alpha_{x;xy} = - \alpha_{y;yy}~. 
\end{align}
\end{subequations}
Thus, ${\mathcal C}_3$ symmetry allows only two independent responses: $\alpha_{x;yy}$ and $\alpha_{y;xx}$. Additionally, each response tensor can be expressed as a combination of a symmetric and an antisymmetric part, in terms of the current and either of the two field indices. These are analogous to the dissipative and dissipationless parts of the second-order charge responses \cite{kamal2023intrinsic}. Interestingly, the above relations force the antisymmetric contributions to vanish in ${\cal C}_3$ symmetric systems. 

The nonlinear anomalous contribution $\alpha^{\rm NA}_{a;bc}$, which originates from the Berry curvature $\Omega^{ab}_n$, is antisymmetric in indices $a$ and $b$. As a result, $\alpha^{\rm NA}_{a;bc}$ does not contribute to NNE or to NSE. This behavior is evident in Fig.~\ref{fig3}(a), where $\alpha^{\rm NA}_{y;xx}$ has a small value. The finite but small value arises from the small ${\cal C}_3$ symmetry breaking of the tight-binding Hamiltonian in Eq.~\eqref{eq:Ham}. 

In contrast, $\alpha^{\rm NSJ}_{a;bc}$ and $\alpha^{\rm NSK}_{a;bc}$ are neither symmetric nor antisymmetric in indices $a$ and $b$, and therefore contribute finite values to both the NNE and NSE. In an ideal $\mathcal{C}_3$ symmetric system, Eq.~\eqref{eq:C3_sym} implies that the nonlinear conductivities associated with the NNE and NSE should be equal in magnitude but opposite in sign. This is also seen in our numerical results in Fig.~\ref{fig3}(b–c) with a slight deviation arising from the marginal breaking of the ${\mathcal C}_3$ symmetry. 

Taken together, our analysis demonstrates that in time-reversal–symmetric systems that are nearly ${\mathcal C}_3$ symmetric, the nonlinear thermoelectric response is dominated by extrinsic scattering mechanisms. The dominant extrinsic channel is determined by competition between the NSJ and NSK processes and ultimately by the system's properties. 

\section{Conclusion}

In this work, we have developed a unified semiclassical theory of the nonlinear Nernst and Seebeck effects that systematically incorporates disorder-mediated asymmetric scattering. Our analysis reveals six distinct nonlinear thermoelectric contributions: four extrinsic channels to the nonlinear Nernst effect in addition to the nonlinear Drude and Berry-curvature–dipole terms, and two extrinsic channels to the nonlinear Seebeck effect beyond the Drude response. Through a detailed symmetry analysis, we established the conditions under which each mechanism is allowed. 

As a concrete case study, we applied our framework to ABA-stacked trilayer graphene, a non-magnetic system recently reported to exhibit giant nonlinear Nernst signals. Our calculations demonstrate that extrinsic skew-scattering mechanisms dominate both the nonlinear Nernst and Seebeck responses, producing sizable nonlinear voltages that lie well within experimental detection capabilities. These results corroborate recent experiments and identify asymmetric impurity scattering as the microscopic origin of the observed giant responses.

Altogether, our work establishes extrinsic nonlinear thermoelectricity as a symmetry- and band-geometry governed phenomenon. Beyond graphene multilayers, the framework presented here provides general design principles for identifying materials in which extrinsic nonlinear thermoelectric responses are naturally enhanced. This establishes a clear route toward efficient, magnetic-field-free thermoelectric and caloritronic functionalities.

\section{Acknowledgment}
H.V. acknowledges the Ministry of Education, Government of India, for financial support through the Prime Minister’s Research Fellowship. A.A. acknowledges funding from the Core Research Grant by ANRF (Sanction No. CRG/2023/007003), Department of Science and Technology, India.

\appendix
\section{Detailed discussion on symmetric and antisymmetric scattering rates}\label{app:assymmetric_SR}
According to the Fermi-Golden rule, the scattering rate of the Bloch electrons is given by Eq.~\eqref{eq:sr_def}, which depends on the wavefunction $\vert \Psi_{l'}\rangle$ of the full Hamiltonian. Using Eq.~\eqref{eq:lip_sch} for weak disorder potential, we expand $\vert \Psi_{l'}\rangle$ in terms of Bloch states as
\bea\label{eq:Lip_Sch} 
&&\ket{\Psi_{l'}} \approx \ket{u_{l'}} + \sum_{l''} \frac{V_{l''l'}}{\e_{l'} - \e_{l''} + i\eta} \ket{u_{l''}}  \\ 
&&+ \sum_{l''} \sum_{l'''} \frac{V_{l''l'''} V_{l'''l'}}{(\e_{l'} - \e_{l''} + i\eta)(\e_{l'} - \e_{l'''} + i\eta)}\ket{u_{l''}} + \cdots~,\nn
\eea 
where $V_{l''l'}  = \langle u_{l''}\vert  V_{\rm imp} \vert u_{l'} \rangle$ is the expectation value of the impurity potential $V_{\rm imp}$ between Bloch states $\ket{u_{l''}}$ and $\ket{u_{l'}}$. Upon inserting this expansion in Eq.~\eqref{eq:sr_def}, the scattering rate can be computed as a power series in the impurity potential: $w_{ll'} = w_{ll'}^{(2)} + w_{ll'}^{(3)} + w_{ll'}^{(4)} + \cdots$. Here, $w_{ll'}^{(\nu)}$ with $\nu \in [2,3,4,\cdots]$ denotes the $\nu$th-order scattering rate depending on the $\nu$th power of the impurity potential. The explicit form of the second, third, and fourth-order scattering rates is given by
\begin{subequations}
\begin{align}
w^{(2)}_{ll'} &= \frac{2\pi}{\hbar} \langle  V_{ll'} V_{l'l} \rangle_{\rm dis}~\delta(\e_l - \e_{l'})~, \\
w^{(3)}_{ll'} &= \frac{2\pi}{\hbar} \sum_{l''} \bigg( \frac{\langle V_{ll''} V_{l''l'} V_{l'l} \rangle_{\rm dis}}{\e_{l'} - \e_{l''} + i\eta} \nn \\ 
& \ +  \frac{\langle V_{ll'} V_{l'l''} V_{l''l} \rangle_{\rm dis}}{\e_{l'} - \e_{l''} - i\eta}\bigg) \delta(\e_l - \e_{l'})~, \\ 
w^{(4)}_{ll'} &= \frac{2\pi}{\hbar} \sum_{l''} \sum_{l'''} \bigg( \frac{\langle V_{ll'''} V_{l'''l'} V_{l'l''} V_{l''l} \rangle_{\rm dis}}{(\e_{l'} - \e_{l''} - i\eta) (\e_{l'} - \e_{l'''} + i\eta)} \nn \\ 
& + \frac{\langle V_{ll''} V_{l''l'''} V_{l'''l'} V_{l'l} \rangle_{\rm dis}}{(\e_{l'} - \e_{l''} + i\eta) (\e_{l'} - \e_{l'''} + i\eta)} \label{eq:4th_sr} \\ 
& + \frac{\langle V_{ll'} V_{l'l'''} V_{l'''l''} V_{l''l} \rangle_{\rm dis}}{(\e_{l'} - \e_{l''} - i\eta) (\e_{l'} - \e_{l'''}  - i\eta)} \bigg) \delta(\e_l - \e_{l'})~. \nn
\end{align}
\end{subequations} 
In the above equations, we consider the impurity potential to be a real quantity and use $V^{*}_{ll'} = V_{l'l}$. Interestingly, the lowest-order scattering rtae $w^{(2)}_{ll'}$ is completely symmetric on interchanging $l$ and $l'$, while $w^{(3)}_{ll'}$ and $w^{(4)}_{ll'}$ contain both symmetric and anti-symmetric parts. Within the weak disorder limit, the symmetric components of $w^{(3,4)}_{ll'}$ can be neglected as they only renormalize the second-order symmetric scattering rate $w^{(2)}_{ll'}$, leading to $w^S_{ll'} \approxeq w^{(2)}_{ll'}$. In contrast, the antisymmetric scattering rate contains contributions from both $w^{(3)}_{ll'}$ and $w^{(4)}_{ll'}$, as they are qualitatively different. For instance, the fourth-order scattering rate $w^{(4)}_{ll'}$ survives even if the third-order scattering rate $w^{(3)}_{ll'}$ vanishes. Therefore, we have an antisymmetric scattering rate as $w^A_{ll'} \approxeq w^{(3), A}_{ll'} + w^{(4), A}_{ll'}$, where $w^{(3), A}_{ll'}$ is the third-order antisymmetric scattering rate, and $w^{(4), A}_{ll'}$ denotes the fourth-order antisymmetric scattering rate. 

We compute the antisymmetric scattering rate by following Eq.~\eqref{eq:sy_nd_as_sr}. Thus, in determining the third-order asymmetric scattering rate, we will encounter a term of the form: $(\e_{l'} - \e_{l''} + i\eta)^{-1} - (\e_{l} - \e_{l''} - i\eta)^{-1}$. Due to the presence of the $\delta(\e_l - \e_{l'})$, we can replace $\e_{l'}$ by $\e_l$ and in $\eta \to 0^+$ limit, this term reduces to
\be\label{eq:lorentzian1}
\lim_{\eta \to 0^+} \left[ \frac{1}{\e_{l} - \e_{l''} + i\eta} - \frac{1}{\e_{l} - \e_{l''} - i\eta} \right] = - i2\pi \delta(\e_{l} - \e_{l''})~.
\ee
In obtaining this result, we have used the Lorentzian representation of the Dirac delta function, $i.e.,~\delta(\e) = \lim_{\eta \to 0} \frac{\eta}{\pi (\e^2 + \eta^2)}$. Consequently, we obtained the third-order antisymmetric scattering rate as 
\bea
&& w^{(3), A}_{ll'} = -\frac{2\pi^2 i}{\hbar} \sum_{l''} \left( \langle V_{ll''} V_{l''l'} V_{l'l} \rangle_{\rm dis} - c.c. \right) \nn \\ 
&& \ \ \ \ \ \ \ \ \ \ \times \delta(\e_l - \e_{l'})\delta(\e_{l} - \e_{l''})~,   \\
&&\equiv \frac{4\pi^2}{\hbar} \sum_{l''} {\rm Im} \langle V_{ll''} V_{l''l'} V_{l'l} \rangle_{\rm dis} \delta(\e_l - \e_{l'})\delta(\e_{l} - \e_{l''})~. \nn 
\eea 
In the above equation, `$c.c.$' and `Im' respectively denote the complex conjugate and Imaginary part of a complex quantity. For the non-degenerate energy bands, the presence of $\delta(\e_l - \e_{l'}) \delta(\e_l - \e_{l''})$ forces composite indices $l,~l',~l''$ to have same band index with different wave vectors. This means $l\equiv (n,\kb)$, $l' \equiv (n,\kb')$, and $l'' \equiv (n,\kb'')$. Thus, we rewrite $w^{(3), A}_{ll'}$ for $n$th band as 
\bea\label{eq:3rd_as_sr}
w^{(3), A}_{n, \kb\kb'} &=& \frac{4\pi^2}{\hbar} \sum_{\kb''} {\rm Im} \langle V^{nn}_{\kb\kb''} V^{nn}_{\kb''\kb'} V^{nn}_{\kb'\kb} \rangle_{\rm dis} \nn \\ 
&& \times \delta(\e^n_\kb - \e^n_{\kb'})\delta(\e^n_{\kb} - \e^n_{\kb''})~.
\eea 

Similarly, to deduce the fourth-order antisymmetric scattering rate, we compute $w^{(4)}_{l'l}$ from Eq.~\eqref{eq:4th_sr} and apply $l'' \leftrightarrow l'''$ as they are dummy indices to make the numerators of $w^{(4)}_{ll'}$ and $w^{(4)}_{l'l}$ similar. For ease of calculation, we assign the following notations:
\begin{subequations}
    \begin{align}
    \langle V_{ll'''} V_{l'''l'} V_{l'l''} V_{l''l} \rangle_{\rm dis} &= \mathcal{V}_1~, \\
    \langle V_{ll''} V_{l''l'''} V_{l'''l'} V_{l'l} \rangle_{\rm dis} &= \mathcal{V}_2~, \\ 
    \langle V_{ll'} V_{l'l'''} V_{l'''l''} V_{l''l} \rangle_{\rm dis} &= \mathcal{V}_2^*~. 
    \end{align}
\end{subequations}
In doing so, we encounter some terms of the following form: $[(\e_1 \pm i\eta)(\e_2 \pm i \eta)]^{-1} - c.c.$ with $\e_1 = \e_{l'} - \e_{l''}$ and $\e_2 = \e_{l'} - \e_{l'''}$ and $\eta \to 0$. Similar to Eq.~\eqref{eq:lorentzian1}, we simplify such terms as 
\begin{subequations}
    \begin{align}
        \lim_{\eta \to 0^+}\left[ \frac{1}{(\e_1 \pm i\eta)(\e_2 \pm i \eta)} - c.c.\right] &\equiv \mp i2\pi \left[ \frac{\delta(\e_1)}{\e_2} + \frac{\delta(\e_2)}{\e_1}\right]~, \nn \\ 
        \lim_{\eta \to 0^+}\left[ \frac{1}{(\e_1 \pm i\eta)(\e_2 \mp i \eta)} - c.c.\right] &\equiv \mp i2\pi \left[ \frac{\delta(\e_1)}{\e_2} - \frac{\delta(\e_2)}{\e_1}\right]~. \nn 
    \end{align}
\end{subequations}
Using all this discussion, we calculate the fourth-order antisymmetric scattering rate as 
\bea 
w^{(4), A}_{ll'} &=& \frac{2\pi^2 i}{\hbar} \sum_{l''} \sum_{l'''} \bigg[ \mathcal{V}_1 \left( \frac{\delta(\e_1)}{\e_2} - \frac{\delta(\e_2)}{\e_1} \right) \\ 
&&- (\mathcal{V}_2 - \mathcal{V}_2^*) \left( \frac{\delta(\e_1)}{\e_2} + \frac{\delta(\e_2)}{\e_1} \right) \bigg]\delta(\e_l - \e_{l'})~. \nn
\eea 
In the first term of the above equation, to take the delta function common, we apply $l'' \leftrightarrow l'''$, which makes $V_1 \to V_1^*$, and the above equation reduces to 
\begin{widetext}
\bea 
&& w^{(4), A}_{ll'} = -\frac{4\pi^2}{\hbar} \sum_{l''} \sum_{l'''} \bigg[ {\rm Im} \mathcal{V}_1 \frac{\delta(\e_{l'} - \e_{l''})}{\e_{l'} - \e_{l'''}} - {\rm Im} \mathcal{V}_2 \left( \frac{\delta(\e_{l'} - \e_{l''})}{\e_{l'} - \e_{l'''}} + \frac{\delta(\e_{l'} - \e_{l'''})}{\e_{l'} - \e_{l''}} \right)  \bigg] \delta(\e_l - \e_{l'})~.  \\
&\equiv& -\frac{4\pi^2}{\hbar} \sum_{l''} \sum_{l'''} \bigg[ {\rm Im} \langle V_{ll'''} V_{l'''l'} V_{l'l''} V_{l''l} \rangle_{\rm dis} \frac{\delta(\e_{l'} - \e_{l''})}{\e_{l'} - \e_{l'''}}  - {\rm Im} \langle V_{ll''} V_{l''l'''} V_{l'''l'} V_{l'l} \rangle_{\rm dis} \left( \frac{\delta(\e_{l'} - \e_{l''})}{\e_{l'} - \e_{l'''}} + \frac{\delta(\e_{l'} - \e_{l'''})}{\e_{l'} - \e_{l''}} \right)  \bigg] \delta(\e_l - \e_{l'})~, \nn  \\ 
&=& -\frac{4\pi^2}{\hbar} \sum_{l''} \sum_{l'''} \bigg[ {\rm Im} \langle V_{ll'''} V_{l'''l'} V_{l'l''} V_{l''l} \rangle_{\rm dis} - {\rm Im} \langle V_{ll''} V_{l''l'''} V_{l'''l'} V_{l'l} \rangle_{\rm dis}  - {\rm Im} \langle V_{ll'''} V_{l'''l''} V_{l''l'} V_{l'l} \rangle_{\rm dis}\bigg] \frac{\delta(\e_{l} - \e_{l'}) \delta(\e_{l'} - \e_{l''})}{\e_{l'} - \e_{l'''}}~. \nn
\eea
\end{widetext}
The last line of the above equation is obtained by applying $l'' \leftrightarrow l'''$ in the $\delta(\e_{l'} - \e_{l'''})$ term. Here, we have to be careful with the summations $\sum_{l''} $ and $\sum_{l'''}$. For the nondegenerate bands, the delta function $\delta(\e_l -\e_{l'})\delta(\e_l -\e_{l''})$  forces $l,~l',~l''$ to have same band index with different wave vectors, while the term $1/(\e_{l'} - \e_{l'''})$ leads to an unphysical divergent when $\e_{l'} = \e_{l'''}$. Thus, to exclude such spurious singularly, we consider $l''' \equiv (n', \kb''')$ with $n' \ne n$. Therefore, the fourth-order antisymmetric scattering rate for the $n$th band is given by
\bea\label{eq:4th_as_sr}
&& w^{(4), A}_{n, \kb \kb'} =  -\frac{4\pi^2}{\hbar} \sum_{\kb''} \sum_{\kb'''} \sum_{n'}^{n' \ne n} \Bigg[ {\rm Im} \langle V^{nn'}_{\kb \kb'''} V^{n'n}_{\kb'''\kb'} V^{nn}_{\kb'\kb''} V^{nn}_{\kb''\kb} \rangle_{\rm dis} ~\nn \\ 
&& \ \ \ \ \  - {\rm Im} \langle V^{nn}_{\kb\kb''} V^{nn'}_{\kb''\kb'''} V^{n'n}_{\kb'''\kb'} V^{nn}_{\kb'\kb} \rangle_{\rm dis} -   \\ 
&& \ \ \ \ \ {\rm Im} \langle V^{nn'}_{\kb\kb'''} V^{n'n}_{\kb'''\kb''} V^{nn}_{\kb''\kb'} V^{nn}_{\kb'\kb} \rangle_{\rm dis} \Bigg] \frac{\delta(\e^n_\kb - \e^n_{\kb'}) \delta(\e^n_{\kb'} - \e^n_{\kb''})}{\e^n_{\kb'} - \e^{n'}_{\kb'''}}~. \nn 
\eea 
%

\section{Explicit connection of side-jump velocity and antisymmetric scattering rate with Berry curvature} \label{app:simplification}
Accurate determination of the side-jump velocity and the antisymmetric scattering rate is essential for capturing the extrinsic mechanisms underlying nonlinear transport phenomena. While closed-form analytical expressions can be derived for simple two-band models, the calculations rapidly become cumbersome and often intractable in multiband systems.

Although the final expressions of both the side-jump velocity and the antisymmetric scattering rate are gauge invariant, their intermediate components remain explicitly gauge dependent. This sensitivity to the gauge choice of Bloch wavefunctions poses a major challenge in numerical simulations, as even minor discontinuities or phase inconsistencies in the Bloch states can lead to spurious or nonphysical results.

To address these issues, we reformulate both quantities in terms of the Berry curvature, a fundamentally gauge-invariant band geometric property of Bloch bands. This reformulation eliminates gauge ambiguities, enabling reliable, consistent numerical results in complex multiband systems.

\subsection{Side-jump velocity}
In this paper, we model disorder impurities as a static, short-range delta-like potential. Thus, the side-jump velocity can be expressed as
\be 
{\bm v}^{\rm sj}_{n}(\kb) = - \frac{2\pi}{\hbar} n_i V_0^2 \sum_{n', \kb'} \vert \bra{u_n^\kb} u^{\kb'}_{n'}\rangle \vert^2 \delta(\e^\kb_n - \e^{\kb'}_{n'}) \delta {\bm r}_{nn'}(\kb,\kb')~.
\ee 
To proceed, we focus on the low-energy regime of the electronic band structure, as the experimentally observable physics arises in this region. 
Within the low-energy limit, we assume that the electronic band structure is non-degenerate and electrons do not undergo interband transitions. Thus, the delta function $\delta(\e^\kb_n - \e^{\kb'}_{n'})$ enforces $n' \to n$, which makes the positional shift vector along the $a$-direction as
\be\label{eq:pos_shift_app}
\delta r^a_{nn}(\kb, \kb') = R^a_{nn}(\kb)  - R^a_{nn}(\kb') - (\partial_a + \partial'_a) {\rm arg}(\bra{u^{\kb}_n} u^{\kb'}_n\rangle)~,
\ee
where $\partial_a = \partial/\partial k_a$, $\partial'_a = \partial/\partial k'_a$. The terms $R^a_{nn}(\kb) = \bra{u^\kb_n}i\partial_a u_n^\kb \rangle$ and $R^a_{nn}(\kb') = \langle u^{\kb'}_n \vert i\partial'_a u_n^{\kb'} \rangle$ denote the intraband Berry connections at $\kb$ and $\kb'$, respectively. To proceed analytically, we consider the weak-scattering limit where the dominant scattering processes involve small momentum transfer ${\bm q} = (\kb' - \kb)  \to 0$. Within this limit, we expand the Bloch state $\vert u^{\kb'}_n\rangle$ around $\kb$ as
\be \label{eq:un_expand}
\ket{u^{\kb'}_n} \approx \ket{u^{\kb}_n} + q_b \partial_b \ket{u^{\kb}_n} + \frac{1}{2} q_b q_c \partial_b \partial_c \ket{u^{\kb}_n}+ \cdots~, 
\ee 
where the Einstein summation convention is employed. With this expansion, it is straightforward to show that \( \partial'_a \vert u^{\kb'}_n \rangle =  \ket{\partial_a u^{\kb}_n} + q_b  \ket{ \partial_a \partial_b u^{\kb}_n}\), which allows us to express Berry connection at $\kb'$ as 
\be 
R^a_{nn}(\kb') = R^a_{nn}(\kb)  + i q_b \bigg[ \bra{u^\kb_n} \partial_a\partial_b u^\kb_n \rangle  + \bra{\partial_b u^\kb_n} \partial_a u^\kb_n \rangle\bigg]~. \nn
\ee 
Using the identity \( \bra{u^\kb_n} \partial_a\partial_b u^\kb_n \rangle = - \bra{\partial_a u^\kb_n} \partial_b u^\kb_n \rangle - i \partial_a R^b_{nn} \) in the above equation, we find
\be 
R^a_{nn}(\kb') = R^a_{nn}(\kb) - q_b \Omega^{ab}_n(\kb) +  q_b \partial_a R^b_{nn}(\kb)~, \nn
\ee 
where \( \Omega^{ab}_n(\kb) = i \left[ \bra{\partial_a u^\kb_n} \partial_b u^\kb_n\rangle -  \bra{\partial_b u^\kb_n} \partial_a u^\kb_n\rangle \right]\) is the Berry curvature tensor~\cite{xiao2010berry}. 
Next, for small $\bm{q}$, the overlap between nearby Bloch states can be approximated as \[ \bra{u^{\kb}_n} u^{\kb'}_n\rangle = 1 - i q_b R^b_{nn}(\kb) \approx e^{-iq_b R^b_{nn}(\kb)} \quad {\rm as} \quad  q_b \to 0~, \] which gives  \[ {\rm arg}(\bra{u^{\kb}_n} u^{\kb'}_n\rangle) = -q_b R^b_{nn}(\kb)\] and \[ (\partial_a + \partial'_a) {\rm arg}(\bra{u^{\kb}_n} u^{\kb'}_n\rangle) = -q_b \partial_a R^b_{nn}(\kb)~.\]

Substituting these results into the expression for the coordinate shift, we obtain 
\bea
\delta r^a_{nn} (\kb, \kb') &=& q_b \Omega^{ab}_n(\kb) = (k'_b - k_b) \epsilon_{abc} \Omega^c_n(\kb)~, \nn \\  
&=& [(\kb' - \kb) \times {\bm \Omega}_n(\kb)]_a~, \nn 
\eea 
where we have used $\Omega^{ab}_n \equiv \epsilon_{abc} \Omega^c_n$ with $\epsilon_{abc}$ being the Levi-Civita tensor. More compactly, the coordinate shift can be expressed as $\delta {\bm r}_{nn'} (\kb, \kb') = (\kb' - \kb) \times {\bm \Omega}_n(\kb)$. 
Since the overlap factor $\vert \bra{u^{\kb}_n} u^{\kb'}_n\rangle\vert^2 \approx 1$ in the long-wavelength limit, the side-jump velocity reduces
\be \label{eq:sj_vel_simp}
{\bm v}^{\rm sj}_n (\kb) = \frac{2\pi}{\hbar} n_i V_0^2 \sum_{\kb'} [ (\kb - \kb') \times {\bm \Omega}_n(\kb) ] \delta(\e^\kb_n -\e^{\kb'}_n )~.
\ee 
This compact expression provides a direct connection between the side-jump velocity and the Berry curvature, thereby offering a physically transparent and numerically robust way to evaluate extrinsic side-jump effects across realistic material systems.

\subsection{Third-order antisymmetric scattering rate}
Similar to the side-jump velocity simplification, we express the third-order antisymmetric scattering rate defined in Eq.~\eqref{eq:3rd_as_sr} as 
\bea\label{eq:3rd_sc_rt_new} 
w^{(3), A}_{n, \kb\kb'} &=& \frac{4\pi^2}{\hbar} n_i V_1^3 \sum_{\kb''} {\rm Im} \bigg[ \langle u^\kb_n \vert u^{\kb''}_n\rangle \langle u^{\kb''}_n \vert u^{\kb'}_n\rangle \langle u^{\kb'}_n \vert u^{\kb}_n\rangle \bigg]~\nn \\ 
&& \times \delta(\e^\kb_n - \e^{\kb'}_n)\delta(\e^\kb_n - \e^{\kb''}_n)~. 
\eea 
Due to the short-range scattering potential, the electron experiences a small momentum transfer during scattering events. Thus, the crystal momentum $\kb'$ and $\kb''$ are very close to $\kb$. Therefore, we define ${\bm q} = (\kb^{'}- \kb) \to 0$ and ${\bm q}' = (\kb^{''}- \kb) \to 0$ to expand the corresponding Bloch wavefunctions around $\kb$. Similar to Eq.~\eqref{eq:un_expand}, we also expand $\vert u^{\kb''}_n \rangle$ around $\kb$ as 
\be 
\vert u^{\kb''}_n \rangle \approx \ket{u^{\kb}_n} + q'_b \partial_b \ket{u^{\kb}_n} + \frac{1}{2} q'_b q'_c \partial_b \partial_c \ket{u^{\kb}_n}+ \cdots~.
\ee 
Using this expansion, the relevant overlaps between neighboring Bloch states are obtained as
\bea\label{eq:overlap}
\langle u^\kb_n \vert u^{\kb''}_n \rangle &=& 1 - i q'_b R^b_{nn} + \frac{1}{2} q'_b q'_c \langle u^\kb_n \vert \partial_b \partial_c u^\kb_n \rangle ~, \\ 
\langle u^{\kb''}_n \vert u^{\kb'}_n \rangle &=& 1 +  i (q'_b - q_b) R^b_{nn} + \frac{1}{2} q_b q_c \langle  u^\kb_n \vert \partial_b \partial_c u^\kb_n \rangle \nn \\
&& + q_b q'_c \langle \partial_c u^\kb_n \vert \partial_b u^\kb_n \rangle + \frac{1}{2} q'_b q'_c \langle  \partial_b \partial_c u^\kb_n \vert  u^\kb_n \rangle~. \nn \\ 
\langle u^{\kb'}_n \vert u^{\kb}_n \rangle &=& 1  +  i q_b R^b_{nn}  + \frac{1}{2} q_b q_c \langle \partial_b \partial_c u^\kb_n \vert  u^\kb_n \rangle~,\nn
\eea 
where we used an identity $\langle \partial_b u^\kb_n \vert u^\kb_n \rangle = - \langle u^\kb_n \vert \partial_b u^\kb_n\rangle$. From hereafter, $R^b_{nn}$ denotes the intraband Berry connection evaluated at $\kb$. The product of these three overlaps up to second order in $\bm{q}$ and $\bm{q}'$ yields
\bea 
&& \langle u^\kb_n \vert u^{\kb''}_n \rangle \langle u^\kb_n  \vert u^{\kb''}_n \rangle \langle u^{\kb'}_n \vert u^{\kb}_n \rangle = 1 +  q_b q'_c \langle \partial_c u^\kb_n \vert \partial_b u^\kb_n \rangle \nn \\ 
&& +  (q_b q_c + q'_b q'_c) {\rm Re}\bigg[ \langle u^\kb_n \vert \partial_b \partial_c u^\kb_n \rangle \bigg]  \nn \\ 
&&+ [ q_b q'_c + (q'_b - q_b)(q'_c - q_c)] R^b_{nn} R^c_{nn}~, \nn 
\eea 
where `Re' stands for the real part of a complex-valued quantity. Among these terms, only the second term carries an imaginary part, since both $\bm{q}$ and ${\bm R}_{nn}$ are real. Hence, the imaginary part of this product is 
\bea 
{\rm Im} \bigg[ \langle u^\kb_n \vert u^{\kb''}_n \rangle \langle u^\kb_n  \vert u^{\kb''}_n \rangle \langle u^{\kb'}_n \vert u^{\kb}_n \rangle \bigg] = - \frac{1}{2} q_b q'_c \Omega^{cb}_n = \frac{1}{2} q_b q'_c \Omega^{bc}_n, \nn 
\eea 
where we have used the definition of the Berry curvature tensor $\Omega^{bc}_n = -2 {\rm Im}\left[ \langle \partial_b u^\kb_n \vert \partial_c u^\kb_n \rangle \right]$ along with its antisymmetric property $\Omega^{cb}_n = -\Omega^{bc}_n$~\cite{Vanderbilt2018}. Using the pseudovector form of Berry curvature,  $\Omega^{bc}_n \equiv \epsilon_{abc} \Omega^a_n $, we can easily show 
\bea 
q_b q_c'  \Omega^{bc}_n &=& \epsilon_{abc} q_b q'_c \Omega^a_n = ({\bm q} \times {\bm q}')_a \Omega^a_n  = ({\bm q} \times {\bm q}') \cdot {\bm \Omega}_n \nn \\ 
&&= [ (\kb' - \kb) \times (\kb'' - \kb) ] \cdot {\bm \Omega}_n~,  \nn \\ 
&& = -[ (\kb \times \kb'')  + (\kb'' \times \kb') + (\kb' \times \kb) ] \cdot {\bm \Omega}_n~. \nn 
\eea 
Thus, the imaginary part can be rewritten as
\bea 
&& {\rm Im} \bigg[ \langle u^\kb_n \vert u^{\kb''}_n \rangle \langle u^\kb_n  \vert u^{\kb''}_n \rangle \langle u^{\kb'}_n \vert u^{\kb}_n \rangle \bigg]  \nn \\ 
&& = - \frac{1}{2} [ (\kb \times \kb'')  + (\kb'' \times \kb') + (\kb' \times \kb) ] \cdot {\bm \Omega}_n~. \nn 
\eea 
Substituting this result into Eq.~\eqref{eq:3rd_sc_rt_new}, the third-order antisymmetric scattering rate becomes
\bea \label{eq:3rd_skr_simp}
&&w^{(3), A}_{n, \kb \kb'} = -\frac{2\pi^2}{\hbar} n_i V_1^3 \sum_{\kb''} [ (\kb \times \kb'')  + (\kb'' \times \kb') + \nn \\ 
&& \quad  (\kb' \times \kb) ] \cdot {\bm \Omega}_n \delta(\e^\kb_n - \e^{\kb'}_n)\delta(\e^\kb_n - \e^{\kb''}_n)~. 
\eea 
This compact gauge-invariant expression establishes a direct link between the third-order antisymmetric scattering rate and the Berry curvature. It provides a clear physical picture that skew scattering is intrinsically governed by the underlying geometric properties of Bloch bands, encapsulated by ${\bm \Omega}_n$. This formulation thus enables a transparent and reliable evaluation of extrinsic nonlinear transport processes in multiband systems.

\subsection{Fourth-order antisymmetric scattering rate}
Following a similar strategy as adopted in the third-order antisymmetric scattering rate, we recast the fourth-order antisymmetric scattering rate using Eq.~\eqref{eq:4th_as_sr} as 
\bea\label{eq:4th_as_rt_new}
&& w^{(4), A}_{n, \kb \kb'} =  -\frac{4\pi^2}{\hbar} n_i^2 V_0^4 \sum_{\kb''} \sum_{\kb'''} \sum_{n'}^{n' \ne n} \frac{\delta(\e_n^\kb - \e_n^{\kb'}) \delta(\e_n^{\kb'} - \e_n^{\kb''})}{\e_n^{\kb} - \e_{n'}^{\kb'''}}~\nn \\
&& \times \Bigg[ {\rm Im} \bigg(\langle u^\kb_n \vert u^{\kb'''}_{n'} \rangle \langle u^{\kb'''}_{n'} \vert u^{\kb'}_n\rangle \langle u^{\kb'}_n \vert u^{\kb''}_n\rangle \langle u^{\kb''}_n \vert u^{\kb}_n\rangle \bigg)~\nn \\ 
&& \quad   - {\rm Im} \bigg(\langle u^\kb_n \vert u^{\kb''}_{n} \rangle \langle u^{\kb''}_{n} \vert u^{\kb'''}_{n'}\rangle \langle u^{\kb'''}_{n'} \vert u^{\kb'}_n\rangle \langle u^{\kb'}_n \vert u^{\kb}_n \rangle \bigg)   \\ 
&& \quad - {\rm Im} \bigg(\langle u^\kb_n \vert u^{\kb'''}_{n'} \rangle \langle u^{\kb'''}_{n'} \vert u^{\kb''}_{n} \rangle \langle u^{\kb''}_{n} \vert u^{\kb'}_n\rangle \langle u^{\kb'}_n \vert u^{\kb}_n \rangle \bigg) \Bigg]  ~. \nn 
\eea 
For small momentum transfer, the crystal momentum $\kb'''$ is very close to $\kb$  or ${\bm q}'' = \kb''' - \kb \to 0$. Expanding the Bloch states to second order in $\bm{q}''$ using Eq.~\eqref{eq:overlap}, we obtain
\bea 
\langle u^\kb_n \vert u^{\kb'''}_{n'} \rangle &=& -i q''_b R^b_{nn'} - \frac{1}{2} q''_b q''_c \langle u^\kb_n \vert \partial_b \partial_c u^{\kb}_{n'}\rangle \nn~, \\
\langle u^{\kb'''}_{n'} \vert u^{\kb'}_n \rangle &=& i (q''_b - q_b) R^b_{n'n} + \frac{1}{2} q_b q_c \langle  u^\kb_{n'} \vert \partial_b \partial_c u^\kb_{n'} \rangle \nn \\
&& + q_b q''_c \langle \partial_c u^\kb_n \vert \partial_b u^\kb_n \rangle + \frac{1}{2} q''_b q''_c \langle  \partial_b \partial_c u^\kb_{n'} \vert  u^\kb_n \rangle~, \nn
\eea 
where ${\bm R}_{nn'} = \langle u^\kb_n \vert i \partial_{\kb} u^{\kb}_{n'} \rangle $ is the interband Berry connection, and we also use orthogonality property of Bloch states $\langle u^\kb_n \vert u^\kb_{n'} \rangle = 0$ for $n' \ne n$. Combining the above expressions with Eq.~\eqref{eq:overlap}, the first term in Eq.~\eqref{eq:4th_as_rt_new} can be evaluated up to second order in $\bm{q}$ as
\bea 
\langle u^\kb_n \vert u^{\kb'''}_{n'} \rangle \langle u^{\kb'''}_{n'} \vert u^{\kb'}_n\rangle \langle u^{\kb'}_n \vert u^{\kb''}_n\rangle \langle u^{\kb''}_n \vert u^{\kb}_n\rangle = q''_b (q''_c - q_c) R^b_{nn'} R^c_{n'n}. \nn
\eea 
The product $R^b_{nn'} R^c_{n'n}$ defines the \textit{quantum geometric tensor}~\cite{watanabe2021chiral, varshney2025intrinsic}, $$Q^{bc}_{nn'} = R^b_{nn'} R^c_{n'n} \equiv {\mathcal G}^{bc}_{nn'} - \frac{i}{2} \Omega^{bc}_{nn'}~,$$ where ${\mathcal G}^{bc}_{nn'}$ and $\Omega^{bc}_{nn'}$ represent the interband quantum metric and Berry curvature tensors, respectively. Hence, the imaginary part of the above product becomes
\bea 
&& {\rm Im} \bigg( \langle u^\kb_n \vert u^{\kb'''}_{n'} \rangle \langle u^{\kb'''}_{n'} \vert u^{\kb'}_n\rangle \langle u^{\kb'}_n \vert u^{\kb''}_n\rangle \langle u^{\kb''}_n \vert u^{\kb}_n\rangle \bigg) \nn \\ 
&& = - \frac{1}{2} q''_b (q''_c - q_c) \Omega^{bc}_{nn'}. \nn 
\eea 
Similarly, the other two terms in Eq.~\eqref{eq:4th_as_rt_new} yield
\bea 
&& {\rm Im} \bigg( \langle u^\kb_n \vert u^{\kb''}_{n} \rangle \langle u^{\kb''}_{n} \vert u^{\kb'''}_{n'}\rangle \langle u^{\kb'''}_{n'} \vert u^{\kb'}_n\rangle \langle u^{\kb'}_n \vert u^{\kb}_n \rangle \bigg) \nn \\ 
&& = - \frac{1}{2} (q''_b - q'_b) (q''_c - q_c) \Omega^{bc}_{nn'}~, \nn \\ 
&& {\rm Im} \bigg( \langle u^\kb_n \vert u^{\kb'''}_{n'} \rangle \langle u^{\kb'''}_{n'} \vert u^{\kb''}_{n} \rangle \langle u^{\kb''}_{n} \vert u^{\kb'}_n\rangle \langle u^{\kb'}_n \vert u^{\kb}_n \rangle \bigg) \nn \\ 
&& = - \frac{1}{2} q''_b (q''_c - q'_c) \Omega^{bc}_{nn'}~. \nn
\eea 

Substituting these results into Eq.~\eqref{eq:4th_as_rt_new} and simplifying, the term inside the square bracket reduces to
\be 
\frac{1}{2} \left[ (\kb \times \kb'') + (\kb'' \times \kb') + (\kb' \times \kb) \right] \cdot {\bm \Omega}_{nn'}~. \nn
\ee 
Next, we evaluate the energy denominator by approximating $\e^\kb_n - \e^{\kb'''}_{n'} \approx (\e^\kb_n - \e^{\kb}_{n'})  + \cdots$ as $\kb''' \to \kb$ in the long-range scattering limit. Hence, 
\be 
\sum_{\kb'''} \frac{1}{\e^\kb_n - \e^{\kb'''}_{n'}} \approx \frac{1}{\e^\kb_n - \e^{\kb}_{n'}}~. \nn 
\ee

Finally, the fourth-order antisymmetric scattering rate becomes
\bea \label{eq:4th_skr_simp}
&& w^{(4), A}_{n, \kb \kb'} =   -\frac{2 \pi^2}{\hbar} n_i^2 V_0^4 \sum_{\kb''} \delta(\e_n^\kb - \e_n^{\kb'}) \delta(\e_n^{\kb'} - \e_n^{\kb''})~\nn \\ 
&& \times  \left[ (\kb \times \kb'') + (\kb'' \times \kb') + (\kb' \times \kb) \right] \cdot \tilde{\bm \Omega}_{n}~, 
\eea 
where we have defined energy normalized Berry curvature $\tilde{\bm \Omega}_n = \sum_{n' \ne n} {\bm \Omega}_{nn'}/(\e^\kb_n - \e^{\kb}_{n'})$. This compact, gauge-invariant expression clearly reveals that the fourth-order antisymmetric scattering rate arises from interband virtual transitions mediated by the Berry curvature. 

Altogether, this direct connection of side-jump velocity, and antisymmetric skew scattering rate with the Berry curvature not only accelerates the numerical computation but also highlights their inherent dependence on band geometric property of the Bloch wavefunctions. 

\subsection{Side-jump velocity and antisymmetric scattering rate for systems with $\mathbf{k}$-even energy dispersion} \label{app:even_E_form_of_ext_quant}
Having derived the compact expressions for the side-jump velocity and the antisymmetric scattering rates for short-ranged delta-function–like static impurities [Eqs.~(\ref{eq:sj_vel_simp}), (\ref{eq:3rd_skr_simp}), and (\ref{eq:4th_skr_simp})], we now consider their specific forms in systems possessing an even energy dispersion under wave-vector inversion, $\e(\kb) = \e(-\kb)$. This condition is naturally satisfied in systems with either inversion symmetry or time-reversal symmetry. For such systems, the following Brillouin-zone summations vanish due to symmetry: 
\be 
\sum_{\kb'} \kb' \delta(\e^\kb_n  - \e^{\kb'}_n) = 0 \quad {\rm and } \quad  \sum_{\kb''} \kb'' \delta(\e^\kb_n  - \e^{\kb''}_n) = 0~. \nn
\ee
Additionally, by invoking the definition of the density of states (DOS), the remaining momentum summations can be expressed as
\bea 
\sum_{\kb'} \delta(\e_n^\kb - \e_n^{\kb'}) = \sum_{\kb''} \delta(\e_n^\kb - \e_n^{\kb''}) =  \mathcal{D}_n(\e_n^\kb)~, \nn 
\eea 
where $\mathcal{D}_n(\e^\kb_n)$ denotes the of the $n$th band for energy $\e^\kb_n$. Substituting these simplifications into the previously obtained general expressions yields the symmetry-reduced forms of the side-jump velocity and antisymmetric scattering rates:
\begin{subequations}
 \begin{align}
     {\bm v}^{\rm sj}_{n, \kb} &= \frac{2\pi}{\hbar} n_i V_0^2 \mathcal{D}_n(\e_n^\kb)  [ \kb \times {\bm \Omega}_n] ~, \\ 
     w^{(3), A}_{n, \kb\kb' } &= \frac{2\pi^2}{\hbar} n_i V_1^3 \mathcal{D}_n(\e^\kb_n) [(\kb \times \kb') \cdot {\bm \Omega}_n]~ \delta(\e^\kb_n  - \e^{\kb'}_n)~, \\ 
     w^{(4), A}_{n, \kb\kb' } &= \frac{2\pi^2}{\hbar} n^2_i V_0^4 \mathcal{D}_n(\e^\kb_n) [(\kb \times \kb') \cdot \tilde{\bm \Omega}_n]~ \delta(\e^\kb_n  - \e^{\kb'}_n).
 \end{align}   
\end{subequations}
These results reveal that, in systems with $\kb$-even energy dispersion, both the side-jump velocity and the antisymmetric scattering rates acquire remarkably simple forms, directly governed by the Berry curvature and the DOS. This connection not only streamlines their practical evaluation but also makes the geometric origin of the extrinsic transport coefficients intuitively clear.

\section{Solving Boltzmann equation for nonequilibrium distribution function}\label{app:sbe_sol}
In this appendix, we will solve Eq.~\eqref{eq:sbe} thoroughly to extract the different contributions of the nonequilibrium distribution function. In this regard, we invoke the relaxation time approximation for the intrinsic elastic collision integral $I^{\rm in}_{el}$, where we approximate it as 
\be\label{eq:rta} 
I^{\rm in}_{el}(f_l) = \sum_{l'} w^{S}_{ll'} (f_l - f_{l'}) = -\frac{f_l - f^0_l}{\tau}~.
\ee 
Here, $\tau$ is the scattering time of the Bloch electrons between two consecutive scatterings, which we consider to be a constant. Further, we recast the side-jump collision integral as: 
\be\label{eq:sj_col_int} 
I^{\rm sj}_{el} (f_l) = -{\bm v}^{\rm sj}_l \cdot \gradient_{\bm r} T \frac{\partial f_l}{ \partial T}~.
\ee 
In obtaining this expression, we approximated $f_{l'} \approx f_l$, which is true due to the presence of $\delta(\e_l - \e_{l'})$ in $w^S_{ll'}$ and used the definition of the side-jump velocity. 

\subsection{Intrinsic distribution function}
We solve Boltzmann Eq.~\eqref{eq:sbe_int} perturbatively by expanding $f_l^{\rm in}$ in different powers of the temperature gradient as $f_l^{\rm in} = f_l^0 + f^{\rm in,(1)}_l  + f^{\rm in,(2)}_l + \cdots$, where the $f^{\rm in,(1)}_l $ and $f^{\rm in,(2)}_l $ respectively denote the first and second-order intrinsic distribution function. Using Eq.~\eqref{eq:rta}, we extract $f^{\rm in, (1)}_l$ and $f^{\rm in, (2)}_l$ from Eq.~\eqref{eq:sbe_int} as 
\bea 
f^{\rm in, (1)}_l &=& -\tau  v^c_l \pdv{f_l^0}{T} \nabla_c T~, \\ 
f^{\rm in, (2)}_l &=& -\tau v^b_l \frac{\partial f_l^{\rm in, (1)}}{ \partial T} \nabla_b T ~, \nn \\
&=& \tau^2 v^b_l v^c_l \pdv[2]{f_l^0}{T} \nabla_b T \nabla_c T~.
\eea 
In calculating these intrinsic distribution functions, we made two assumptions. First, we omit ${\bm v}^{\rm sj}_l$ term from $\dot{\bm r}$ as intrinsic distribution functions should solely depend on intrinsic mechanisms. Second, we consider a constant temperature gradient. In these equations, indices $b,~c$ denote the Cartesian coordinates, and the Einstein summation convention on repeated indices is employed. 

\subsection{Side-jump distribution function}
Following the perturbative approach, we perturbatively expand side-jump distribution function up to second-order as $f^{\rm sj}_l$ as $f^{\rm sj}_l = f^{\rm sj, (1)}_l + f^{\rm sj, (2)}_l + \cdots$. Using Eqs.~(\ref{eq:sbe_sj} \& \ref{eq:sj_col_int}), the Boltzmann equation for the first-order distribution function becomes 
\bea
0 &=& I^{\rm in}_{el}(f^{\rm sj, (1)}_l) + I^{\rm sj}_{el} (f^0_l)~, \nn \\
f^{\rm sj, (1)}_l &=& -\tau v^{{\rm sj},c}_l \pdv{f^0_l}{T} \nabla_c T~.
\eea 
Likewise, the Boltzmann equation for the second-order side-jump distribution function can be written as 
\bea 
&& \dot{\bm r}\cdot \gradient_{\bm r}T \pdv{f^{\rm sj,(1)}}{T} = I^{\rm in}_{el}(f^{\rm sj, (2)}_l) + I^{\rm sj}_{el} (f^{\rm in, (1)}_l)~, \nn \\
 && (v^b_l + v^{{\rm sj}, b}_l) \nabla_b T \pdv{}{T}\left[ -\tau v^{{\rm sj},c}_l \pdv{f^0_l}{T} \nabla_c T \right] = -\frac{f^{\rm sj, (2)}_l}{\tau}  \nn \\ 
 && - v^{ {\rm sj},b}_l \nabla_b T \pdv{}{T}\left[ -\tau  v^c_l \pdv{f_l^0}{T} \nabla_c T \right]~. \nn
\eea 
After simplifying this, we obtain the second-order side-jump distribution function as 
\be
f^{\rm sj, (2)}_l = \tau^2 \left[ (v^b_l v^{{\rm sj},c}_l - v^{{\rm sj},b}_l v^c_l) + v^{{\rm sj},b}_l v^{{\rm sj},c}_l\right] \pdv[2]{f^0_l}{T} \nabla_b T \nabla_c T
\ee
%

\subsection{Skew-scattering distribution function}
Similar to the intrinsic and side-jump distribution function, we expand the skew-scattering distribution function in order of temperature gradient as $f^{\rm sk}_l = f^{\rm sk, (1)}_l + f^{\rm sk, (2)}_l + \cdots  $. Here, we emphasize that in computing $f^{\rm sk}_l$, we neglect the side-jump velocity induced contribution, as side-jump and skew-scattering both are independent quantities. Thus, from Eq.~\eqref{eq:sbe_sk}, we deduce first-order skew-scattering distribution function as 
\bea 
0 &=& I^{\rm in}_{el} (f^{\rm sk, (1)}_l) + I^{\rm sk}_{el} (f^{\rm in, (1)}_l)~,\nn \\ 
f^{\rm sk, (1)}_l &=& -\tau^2 \sum_{l'} w^A_{ll'} \left[ v^c_l \pdv{f_l^0}{T} + v^c_{l'} \pdv{f_{l'}^0}{T} \right] \nabla_c T~. 
\eea 
Likewise, for the second-order skew-scattering distribution function, Eq.~\eqref{eq:sbe_sk} reduces to 
\begin{align}
&v^b_l \nabla_b T \pdv{f_l^{\rm sk, (1)}}{T} = I^{\rm in}_{el} (f^{\rm sk, (2)}_l) + I^{\rm sk}_{el} (f^{\rm in, (2)}_l)~ \\ 
&f^{\rm sk, (2)}_l = \tau^3 \sum_{l'} w^A_{ll'}\left[ 2 v^b_l v^c_l \pdv[2]{f^0_l}{T} + (v^b_l + v^b_{l'}) v^c_{l'} \pdv[2]{f^0_{l'}}{T}\right] \nn \\ 
& \quad \quad \quad \times \nabla_b T \nabla_c T~.\nn 
\end{align}
%

\section{Dimensional analysis of nonlinear thermoelectric conductivities}\label{app:dimension}
Dimensional analysis of the nonlinear thermoelectric conductivities is useful to uncover their explicit dependence on the disorder strength. For this, we express the nonlinear thermoelectric conductivities in units of $\alpha_0 = ea/\tau T^2$. This yields the following dimensionally scaled expressions, 
\begin{align*}
    \alpha^{\rm ND} &= \alpha_0 ~N^3_\tau ~\tilde{\alpha}^{\rm ND}~, \\ 
    \alpha^{\rm NA} &= \alpha_0 ~N^2_\tau~ \tilde{\alpha}^{\rm NA}~,\\ 
    \alpha^{\rm NSJ} &= \alpha_0 ~N^3_\tau ~N_{\rm dis} ~\tilde{\alpha}^{\rm NSJ}~, \\
    \alpha^{\rm NASJ} &= \alpha_0 ~N^2_\tau ~N_{\rm dis}~\tilde{\alpha}^{\rm NASJ}~, \\ 
    \alpha^{\rm NSK3} &= \alpha_0 ~N^4_\tau~ {\tilde N}_{\rm dis}~ \tilde{\alpha}^{\rm NASK3}~, \\ 
    \alpha^{\rm NASK3} &= \alpha_0 ~N^3_\tau~ {\tilde N}_{\rm dis}~ \tilde{\alpha}^{\rm NSK3}~, \nn \\
    \alpha^{\rm NSK4} &= \alpha_0 ~N^4_\tau ~ N^2_{\rm dis}~ \tilde{\alpha}^{\rm NSK4}~, \\ 
    \alpha^{\rm NASK4} &= \alpha_0 ~N^3_\tau~ { N}^2_{\rm dis} ~\tilde{\alpha}^{\rm NASK4}~.
\end{align*}
Here, $\tilde{\alpha}$ denotes the dimensionless nonlinear thermoelectric conductivity associated with each channel. The prefactors $N_\tau$, $N_{\rm dis}$, and ${\tilde N}_{\rm dis}$ are dimensionless parameters that encode the dependence on the symmetric scattering time, Gaussian and non-Gaussian impurity potential component, respectively. These quantities are defined as
\be 
N_\tau = \frac{\tau E_s}{\hbar}, \quad N_{\rm dis} = \frac{n_i V^2_0}{E_s^2 a^2}, \quad  {\tilde N}_{\rm dis} = \frac{n_i V^3_1}{E_s^3 a^4}~, \nn
\ee 
where $E_s$ and $a$ represent the characteristic energy and length scales of the system. In the paper, we take $E_s = 1~{\rm eV}$ and $a = 2.46~{\rm \AA}$, corresponding to the lattice constant of TLG.

\bibliography{Ref}

@article{Sinha_2023,
	author = {Sinha, Subhajit and Adak, Pratap Chandra and Chakraborty, Atasi and Das, Kamal and Debnath, Koyendrila and Sangani, L. D. Varma and Watanabe, Kenji and Taniguchi, Takashi and Waghmare, Umesh V. and Agarwal, Amit and Deshmukh, Mandar M.},
	date = {2022/07/01},
	doi = {10.1038/s41567-022-01606-y},
	id = {Sinha2022},
	isbn = {1745-2481},
	journal = {Nature Physics},
	number = {7},
	pages = {765--770},
	title = {Berry curvature dipole senses topological transition in a moir{\'e}superlattice},
	url = {https://doi.org/10.1038/s41567-022-01606-y},
	volume = {18},
	year = {2022}}

@article{Adak_2024,
	date = {2024/07/01},
	doi = {10.1038/s41578-024-00671-4},
	id = {Adak2024},
	isbn = {2058-8437},
	journal = {Nature Reviews Materials},
	number = {7},
	pages = {481--498},
	title = {Tunable moir{\'e}materials for probing Berry physics and topology},
	url = {https://doi.org/10.1038/s41578-024-00671-4},
	volume = {9},
	year = {2024}}

@article{Chakraborty_2022,
doi = {10.1088/2053-1583/ac8b93},
url = {https://doi.org/10.1088/2053-1583/ac8b93},
year = {2022},
month = {sep},
publisher = {IOP Publishing},
volume = {9},
number = {4},
pages = {045020},
author = {Chakraborty, Atasi and Das, Kamal and Sinha, Subhajit and Adak, Pratap Chandra and Deshmukh, Mandar M and Agarwal, Amit},
title = {Nonlinear anomalous Hall effects probe topological phase-transitions in twisted double bilayer graphene},
journal = {2D Materials},
}

@article{PhysRevB.102.205414,
  title = {Magnus Nernst and thermal Hall effect},
  author = {Mandal, Debottam and Das, Kamal and Agarwal, Amit},
  journal = {Phys. Rev. B},
  volume = {102},
  issue = {20},
  pages = {205414},
  numpages = {10},
  year = {2020},
  month = {Nov},
  publisher = {American Physical Society},
  doi = {10.1103/PhysRevB.102.205414},
  url = {https://link.aps.org/doi/10.1103/PhysRevB.102.205414}
}

@article{Burgos_Atencia31122024,
author = {Rhonald Burgos Atencia and Amit Agarwal and Dimitrie Culcer},
title = {Orbital angular momentum of Bloch electrons: equilibrium formulation, magneto-electric phenomena, and the orbital Hall effect},
journal = {Advances in Physics: X},
volume = {9},
number = {1},
pages = {2371972},
year = {2024},
publisher = {Taylor \& Francis},
doi = {10.1080/23746149.2024.2371972},
URL = {     
        https://doi.org/10.1080/23746149.2024.2371972
}}

@article{PhysRevResearch.2.013088,
  title = {Thermal and gravitational chiral anomaly induced magneto-transport in Weyl semimetals},
  author = {Das, Kamal and Agarwal, Amit},
  journal = {Phys. Rev. Res.},
  volume = {2},
  issue = {1},
  pages = {013088},
  numpages = {9},
  year = {2020},
  month = {Jan},
  publisher = {American Physical Society},
  doi = {10.1103/PhysRevResearch.2.013088},
  url = {https://link.aps.org/doi/10.1103/PhysRevResearch.2.013088}
}

@book{ashcroft1976ssp,
  author = {Ashcroft, N. W. and Mermin, N. D.},
  publisher = {Holt-Saunders},
  title = {{S}olid {S}tate {P}hysics},
  year = 1976
}

@article{disalvo1999thermoelectric,
author = {Francis J. DiSalvo },
title = {Thermoelectric Cooling and Power Generation},
journal = {Science},
volume = {285},
number = {5428},
pages = {703-706},
year = {1999},
doi = {10.1126/science.285.5428.703},
URL = {https://www.science.org/doi/abs/10.1126/science.285.5428.703}}

@article{partoens2006from,
  title = {From graphene to graphite: Electronic structure around the $K$ point},
  author = {Partoens, B. and Peeters, F. M.},
  journal = {Phys. Rev. B},
  volume = {74},
  issue = {7},
  pages = {075404},
  numpages = {11},
  year = {2006},
  month = {Aug},
  publisher = {American Physical Society},
  doi = {10.1103/PhysRevB.74.075404},
  url = {https://link.aps.org/doi/10.1103/PhysRevB.74.075404}
}

@article{sinitsyn2006coordinate,
  title = {Coordinate shift in the semiclassical Boltzmann equation and the anomalous Hall effect},
  author = {Sinitsyn, N. A. and Niu, Q. and MacDonald, A. H.},
  journal = {Phys. Rev. B},
  volume = {73},
  issue = {7},
  pages = {075318},
  numpages = {6},
  year = {2006},
  month = {Feb},
  publisher = {American Physical Society},
  doi = {10.1103/PhysRevB.73.075318},
  url = {https://link.aps.org/doi/10.1103/PhysRevB.73.075318}
}

@article{xiao2007valley,
  title = {Valley-Contrasting Physics in Graphene: Magnetic Moment and Topological Transport},
  author = {Xiao, Di and Yao, Wang and Niu, Qian},
  journal = {Phys. Rev. Lett.},
  volume = {99},
  issue = {23},
  pages = {236809},
  numpages = {4},
  year = {2007},
  month = {Dec},
  publisher = {American Physical Society},
  doi = {10.1103/PhysRevLett.99.236809},
  url = {https://link.aps.org/doi/10.1103/PhysRevLett.99.236809}
}

@Article{uchida2008observation,
author={Uchida, K.
and Takahashi, S.
and Harii, K.
and Ieda, J.
and Koshibae, W.
and Ando, K.
and Maekawa, S.
and Saitoh, E.},
title={Observation of the spin Seebeck effect},
journal={Nature},
year={2008},
month={Oct},
day={01},
volume={455},
number={7214},
pages={778-781},
issn={1476-4687},
doi={10.1038/nature07321},
url={https://doi.org/10.1038/nature07321}
}

@Inbook{Snyder2009thermoelectric,
author="Snyder, G. Jeffrey",
editor="Priya, Shashank
and Inman, Daniel J.",
title="Thermoelectric Energy Harvesting",
bookTitle="Energy Harvesting Technologies",
year="2009",
publisher="Springer US",
address="Boston, MA",
pages="325--336",
isbn="978-0-387-76464-1",
doi="10.1007/978-0-387-76464-1_11",
url="https://doi.org/10.1007/978-0-387-76464-1_11"
}

@article{checklesky2009thermpower,
  title = {Thermopower and Nernst effect in graphene in a magnetic field},
  author = {Checkelsky, Joseph G. and Ong, N. P.},
  journal = {Phys. Rev. B},
  volume = {80},
  issue = {8},
  pages = {081413},
  numpages = {4},
  year = {2009},
  month = {Aug},
  publisher = {American Physical Society},
  doi = {10.1103/PhysRevB.80.081413},
  url = {https://link.aps.org/doi/10.1103/PhysRevB.80.081413}
}

@article{xiao2010berry,
  title = {Berry phase effects on electronic properties},
  author = {Xiao, Di and Chang, Ming-Che and Niu, Qian},
  journal = {Rev. Mod. Phys.},
  volume = {82},
  issue = {3},
  pages = {1959--2007},
  numpages = {0},
  year = {2010},
  month = {Jul},
  publisher = {American Physical Society},
  doi = {10.1103/RevModPhys.82.1959},
  url = {https://link.aps.org/doi/10.1103/RevModPhys.82.1959}
}

@article{qin2011energy,
  title = {Energy Magnetization and the Thermal Hall Effect},
  author = {Qin, Tao and Niu, Qian and Shi, Junren},
  journal = {Phys. Rev. Lett.},
  volume = {107},
  issue = {23},
  pages = {236601},
  numpages = {5},
  year = {2011},
  month = {Nov},
  publisher = {American Physical Society},
  doi = {10.1103/PhysRevLett.107.236601},
  url = {https://link.aps.org/doi/10.1103/PhysRevLett.107.236601}
}

@book{tan2011sustainable,
  title={Sustainable Energy Harvesting Technologies: Past, Present and Future},
  author={Tan, Y.K.},
  isbn={9789533074382},
  url={https://books.google.co.in/books?id=D3WfDwAAQBAJ},
  year={2011},
  publisher={IntechOpen}
}

@article{serbyn2013new,
  title = {New Dirac points and multiple Landau level crossings in biased trilayer graphene},
  author = {Serbyn, Maksym and Abanin, Dmitry A.},
  journal = {Phys. Rev. B},
  volume = {87},
  issue = {11},
  pages = {115422},
  numpages = {10},
  year = {2013},
  month = {Mar},
  publisher = {American Physical Society},
  doi = {10.1103/PhysRevB.87.115422},
  url = {https://link.aps.org/doi/10.1103/PhysRevB.87.115422}
}

@article{koumoto2013thermoelectric,
author = {Koumoto, Kunihito and Funahashi, Ryoji and Guilmeau, Emmanuel and Miyazaki, Yuzuru and Weidenkaff, Anke and Wang, Yifeng and Wan, Chunlei},
title = {Thermoelectric Ceramics for Energy Harvesting},
journal = {Journal of the American Ceramic Society},
volume = {96},
number = {1},
pages = {1-23},
doi = {https://doi.org/10.1111/jace.12076},
url = {https://ceramics.onlinelibrary.wiley.com/doi/abs/10.1111/jace.12076},
year = {2013}
}

@article{sothmann2014thermoelectric,
doi = {10.1088/0957-4484/26/3/032001},
url = {https://dx.doi.org/10.1088/0957-4484/26/3/032001},
year = {2014},
month = {dec},
publisher = {IOP Publishing},
volume = {26},
number = {3},
pages = {032001},
author = {Björn Sothmann and Rafael Sánchez and Andrew N Jordan},
title = {Thermoelectric energy harvesting with quantum dots},
journal = {Nanotechnology}
}

@article{chhatrasal2016,
title = {Recent advances in thermoelectric materials},
journal = {Progress in Materials Science},
volume = {83},
pages = {330-382},
year = {2016},
issn = {0079-6425},
doi = {https://doi.org/10.1016/j.pmatsci.2016.07.002},
url = {https://www.sciencedirect.com/science/article/pii/S0079642516300317},
author = {Chhatrasal Gayner and Kamal K. Kar}
}

@article{kamran2016nernst,
doi = {10.1088/0034-4885/79/4/046502},
url = {https://dx.doi.org/10.1088/0034-4885/79/4/046502},
year = {2016},
month = {mar},
publisher = {IOP Publishing},
volume = {79},
number = {4},
pages = {046502},
author = {Kamran Behnia and Hervé Aubin},
title = {Nernst effect in metals and superconductors: a review of concepts and experiments},
journal = {Reports on Progress in Physics}
}

@book{sakurai2017QM, 
place={Cambridge}, 
edition={2}, 
title={Modern Quantum Mechanics}, 
publisher={Cambridge University Press}, 
author={Sakurai, J. J. and Napolitano, Jim}, 
year={2017}}

@Article{ikhlas2017large,
author={Ikhlas, Muhammad
and Tomita, Takahiro
and Koretsune, Takashi
and Suzuki, Michi-To
and Nishio-Hamane, Daisuke
and Arita, Ryotaro
and Otani, Yoshichika
and Nakatsuji, Satoru},
title={Large anomalous Nernst effect at room temperature in a chiral antiferromagnet},
journal={Nature Physics},
year={2017},
month={Nov},
day={01},
volume={13},
number={11},
pages={1085-1090},
issn={1745-2481},
doi={10.1038/nphys4181},
url={https://doi.org/10.1038/nphys4181}
}

@Article{Rana2018thermopower,
author={Rana, K. Gaurav
and Dejene, Fasil K.
and Kumar, Neeraj
and Rajamathi, Catherine R.
and Sklarek, Kornelia
and Felser, Claudia
and Parkin, Stuart S. P.},
title={Thermopower and Unconventional Nernst Effect in the Predicted Type-II Weyl Semimetal WTe2},
journal={Nano Letters},
year={2018},
month={Oct},
day={10},
publisher={American Chemical Society},
volume={18},
number={10},
pages={6591-6596},
issn={1530-6984},
doi={10.1021/acs.nanolett.8b03212},
url={https://doi.org/10.1021/acs.nanolett.8b03212}
}

@article{Sakai2018giant,
  title = {Giant anomalous Nernst effect and quantum-critical scaling in a ferromagnetic semimetal},
  volume = {14},
  ISSN = {1745-2481},
  url = {http://dx.doi.org/10.1038/s41567-018-0225-6},
  DOI = {10.1038/s41567-018-0225-6},
  number = {11},
  journal = {Nature Physics},
  publisher = {Springer Science and Business Media LLC},
  author = {Sakai,  Akito and Mizuta,  Yo Pierre and Nugroho,  Agustinus Agung and Sihombing,  Rombang and Koretsune,  Takashi and Suzuki,  Michi-To and Takemori,  Nayuta and Ishii,  Rieko and Nishio-Hamane,  Daisuke and Arita,  Ryotaro and Goswami,  Pallab and Nakatsuji,  Satoru},
  year = {2018},
  month = jul,
  pages = {1119–1124}
}

@book{Vanderbilt2018,
  title = {Berry Phases in Electronic Structure Theory: Electric Polarization,  Orbital Magnetization and Topological Insulators},
  ISBN = {9781107157651},
  url = {http://dx.doi.org/10.1017/9781316662205},
  DOI = {10.1017/9781316662205},
  publisher = {Cambridge University Press},
  author = {Vanderbilt,  David},
  year = {2018},
  month = oct 
}

@article{du2019disorder,
  title = {Disorder-induced nonlinear Hall effect with time-reversal symmetry},
  volume = {10},
  ISSN = {2041-1723},
  url = {http://dx.doi.org/10.1038/s41467-019-10941-3},
  DOI = {10.1038/s41467-019-10941-3},
  number = {1},
  journal = {Nature Communications},
  publisher = {Springer Science and Business Media LLC},
  author = {Du,  Z. Z. and Wang,  C. M. and Li,  Shuai and Lu,  Hai-Zhou and Xie,  X. C.},
  year = {2019},
  month = jul 
}

@article{zeng2019nonlinear,
  title = {Nonlinear Nernst effect in bilayer ${\mathrm{WTe}}_{2}$},
  author = {Zeng, Chuanchang and Nandy, Snehasish and Taraphder, A. and Tewari, Sumanta},
  journal = {Phys. Rev. B},
  volume = {100},
  issue = {24},
  pages = {245102},
  numpages = {12},
  year = {2019},
  month = {Dec},
  publisher = {American Physical Society},
  doi = {10.1103/PhysRevB.100.245102},
  url = {https://link.aps.org/doi/10.1103/PhysRevB.100.245102}
}

@article{yu2019topological,
  title = {Topological nonlinear anomalous Nernst effect in strained transition metal dichalcogenides},
  author = {Yu, Xiao-Qin and Zhu, Zhen-Gang and You, Jhih-Shih and Low, Tony and Su, Gang},
  journal = {Phys. Rev. B},
  volume = {99},
  issue = {20},
  pages = {201410},
  numpages = {5},
  year = {2019},
  month = {May},
  publisher = {American Physical Society},
  doi = {10.1103/PhysRevB.99.201410},
  url = {https://link.aps.org/doi/10.1103/PhysRevB.99.201410}
}

@article{kamal2019berry,
  title = {Berry curvature induced thermopower in type-I and type-II Weyl semimetals},
  author = {Das, Kamal and Agarwal, Amit},
  journal = {Phys. Rev. B},
  volume = {100},
  issue = {8},
  pages = {085406},
  numpages = {14},
  year = {2019},
  month = {Aug},
  publisher = {American Physical Society},
  doi = {10.1103/PhysRevB.100.085406},
  url = {https://link.aps.org/doi/10.1103/PhysRevB.100.085406}
}

@article{guin2019anomalous,
  title = {Anomalous Nernst effect beyond the magnetization scaling relation in the ferromagnetic Heusler compound Co2MnGa},
  volume = {11},
  ISSN = {1884-4057},
  url = {http://dx.doi.org/10.1038/s41427-019-0116-z},
  DOI = {10.1038/s41427-019-0116-z},
  number = {1},
  journal = {NPG Asia Materials},
  publisher = {Springer Science and Business Media LLC},
  author = {Guin,  Satya N. and Manna,  Kaustuv and Noky,  Jonathan and Watzman,  Sarah J. and Fu,  Chenguang and Kumar,  Nitesh and Schnelle,  Walter and Shekhar,  Chandra and Sun,  Yan and Gooth,  Johannes and Felser,  Claudia},
  year = {2019},
  month = apr 
}

@article{xu2019large,
  title = {Large Anomalous Nernst Effect in a van der Waals Ferromagnet Fe3GeTe2},
  volume = {19},
  ISSN = {1530-6992},
  url = {http://dx.doi.org/10.1021/acs.nanolett.9b03739},
  DOI = {10.1021/acs.nanolett.9b03739},
  number = {11},
  journal = {Nano Letters},
  publisher = {American Chemical Society (ACS)},
  author = {Xu,  Jinsong and Phelan,  W. Adam and Chien,  Chia-Ling},
  year = {2019},
  month = oct,
  pages = {8250–8254}
}

@article{amin2020review,
title = {Review of wearable thermoelectric energy harvesting: From body temperature to electronic systems},
journal = {Applied Energy},
volume = {258},
pages = {114069},
year = {2020},
issn = {0306-2619},
doi = {https://doi.org/10.1016/j.apenergy.2019.114069},
url = {https://www.sciencedirect.com/science/article/pii/S0306261919317568},
author = {Amin Nozariasbmarz and Henry Collins and Kelvin Dsouza and Mobarak Hossain Polash and Mahshid Hosseini and Melissa Hyland and Jie Liu and Abhishek Malhotra and Francisco Matos Ortiz and Farzad Mohaddes and Viswanath Padmanabhan Ramesh and Yasaman Sargolzaeiaval and Nicholas Snouwaert and Mehmet C. Özturk and Daryoosh Vashaee},
}

@article{Sakai2020Nature,
	author = {Sakai, Akito and Minami, Susumu and Koretsune, Takashi and Chen, Taishi and Higo, Tomoya and Wang, Yangming and Nomoto, Takuya and Hirayama, Motoaki and Miwa, Shinji and Nishio-Hamane, Daisuke and Ishii, Fumiyuki and Arita, Ryotaro and Nakatsuji, Satoru},
	date = {2020/05/01},
	doi = {10.1038/s41586-020-2230-z},
	id = {Sakai2020},
	isbn = {1476-4687},
	journal = {Nature},
	number = {7806},
	pages = {53--57},
	title = {Iron-based binary ferromagnets for transverse thermoelectric conversion},
	url = {https://doi.org/10.1038/s41586-020-2230-z},
	volume = {581},
	year = {2020}}

@article{marchegiani2020nonlinear,
  title = {Nonlinear Thermoelectricity with Electron-Hole Symmetric Systems},
  author = {Marchegiani, G. and Braggio, A. and Giazotto, F.},
  journal = {Phys. Rev. Lett.},
  volume = {124},
  issue = {10},
  pages = {106801},
  numpages = {6},
  year = {2020},
  month = {Mar},
  publisher = {American Physical Society},
  doi = {10.1103/PhysRevLett.124.106801},
  url = {https://link.aps.org/doi/10.1103/PhysRevLett.124.106801}
}

@article{Yang2020giant,
  title = {Giant anomalous Nernst effect in the magnetic Weyl semimetal ${\mathrm{Co}}_{3}{\mathrm{Sn}}_{2}{\mathrm{S}}_{2}$},
  author = {Yang, Haiyang and You, Wei and Wang, Jialu and Huang, Junwu and Xi, Chuanying and Xu, Xiaofeng and Cao, Chao and Tian, Mingliang and Xu, Zhu-An and Dai, Jianhui and Li, Yuke},
  journal = {Phys. Rev. Mater.},
  volume = {4},
  issue = {2},
  pages = {024202},
  numpages = {7},
  year = {2020},
  month = {Feb},
  publisher = {American Physical Society},
  doi = {10.1103/PhysRevMaterials.4.024202},
  url = {https://link.aps.org/doi/10.1103/PhysRevMaterials.4.024202}
}

@article{takao2021,
author = {Takao Mori and Antoine Maignan},
title = {Thermoelectric materials developments: past, present, and future},
journal = {Science and Technology of Advanced Materials},
volume = {22},
number = {1},
pages = {998--999},
year = {2021},
publisher = {Taylor \& Francis},
doi = {10.1080/14686996.2021.1966242},
URL = { https://doi.org/10.1080/14686996.2021.1966242
}
}

@article{papaj2021enhanced,
  title = {Enhanced anomalous Nernst effect in disordered Dirac and Weyl materials},
  author = {Papaj, Micha\l{} and Fu, Liang},
  journal = {Phys. Rev. B},
  volume = {103},
  issue = {7},
  pages = {075424},
  numpages = {12},
  year = {2021},
  month = {Feb},
  publisher = {American Physical Society},
  doi = {10.1103/PhysRevB.103.075424},
  url = {https://link.aps.org/doi/10.1103/PhysRevB.103.075424}
}

@article{kamal2021intrinsic,
  title = {Intrinsic Hall conductivities induced by the orbital magnetic moment},
  author = {Das, Kamal and Agarwal, Amit},
  journal = {Phys. Rev. B},
  volume = {103},
  issue = {12},
  pages = {125432},
  numpages = {10},
  year = {2021},
  month = {Mar},
  publisher = {American Physical Society},
  doi = {10.1103/PhysRevB.103.125432},
  url = {https://link.aps.org/doi/10.1103/PhysRevB.103.125432}
}

@article{Uchida2021transverse,
  title = {Transverse thermoelectric generation using magnetic materials},
  volume = {118},
  ISSN = {1077-3118},
  url = {http://dx.doi.org/10.1063/5.0046877},
  DOI = {10.1063/5.0046877},
  number = {14},
  journal = {Applied Physics Letters},
  publisher = {AIP Publishing},
  author = {Uchida,  Ken-ichi and Zhou,  Weinan and Sakuraba,  Yuya},
  year = {2021},
  month = apr 
}

@article{Pan2021giant,
  title = {Giant anomalous Nernst signal in the antiferromagnet YbMnBi2},
  volume = {21},
  ISSN = {1476-4660},
  url = {http://dx.doi.org/10.1038/s41563-021-01149-2},
  DOI = {10.1038/s41563-021-01149-2},
  number = {2},
  journal = {Nature Materials},
  publisher = {Springer Science and Business Media LLC},
  author = {Pan,  Yu and Le,  Congcong and He,  Bin and Watzman,  Sarah J. and Yao,  Mengyu and Gooth,  Johannes and Heremans,  Joseph P. and Sun,  Yan and Felser,  Claudia},
  year = {2021},
  month = nov,
  pages = {203–209}
}

@article{Wu2021nonlinear,
  title = {Nonlinear anomalous Nernst effect in strained graphene induced by trigonal warping},
  author = {Wu, Ying-Li and Zhu, Ge-Hui and Yu, Xiao-Qin},
  journal = {Phys. Rev. B},
  volume = {104},
  issue = {19},
  pages = {195427},
  numpages = {8},
  year = {2021},
  month = {Nov},
  publisher = {American Physical Society},
  doi = {10.1103/PhysRevB.104.195427},
  url = {https://link.aps.org/doi/10.1103/PhysRevB.104.195427}
}

@article{watanabe2021chiral,
  title = {Chiral Photocurrent in Parity-Violating Magnet and Enhanced Response in Topological Antiferromagnet},
  author = {Watanabe, Hikaru and Yanase, Youichi},
  journal = {Phys. Rev. X},
  volume = {11},
  issue = {1},
  pages = {011001},
  numpages = {30},
  year = {2021},
  month = {Jan},
  publisher = {American Physical Society},
  doi = {10.1103/PhysRevX.11.011001},
  url = {https://link.aps.org/doi/10.1103/PhysRevX.11.011001}
}

@article{zhou2022fundamental,
  title = {Fundamental distinction between intrinsic and extrinsic nonlinear thermal Hall effects},
  author = {Zhou, Da-Kun and Zhang, Zhi-Fan and Yu, Xiao-Qin and Zhu, Zhen-Gang and Su, Gang},
  journal = {Phys. Rev. B},
  volume = {105},
  issue = {20},
  pages = {L201103},
  numpages = {5},
  year = {2022},
  month = {May},
  publisher = {American Physical Society},
  doi = {10.1103/PhysRevB.105.L201103},
  url = {https://link.aps.org/doi/10.1103/PhysRevB.105.L201103}
}

@article{Uchida2022thermoelectrics,
  title = {Thermoelectrics: From longitudinal to transverse},
  volume = {6},
  ISSN = {2542-4351},
  url = {http://dx.doi.org/10.1016/j.joule.2022.08.016},
  DOI = {10.1016/j.joule.2022.08.016},
  number = {10},
  journal = {Joule},
  publisher = {Elsevier BV},
  author = {Uchida,  Ken-ichi and Heremans,  Joseph P.},
  year = {2022},
  month = oct,
  pages = {2240–2245}
}

@article{ma2023anomalous,
  title = {Anomalous Skew-Scattering Nonlinear Hall Effect and Chiral Photocurrents in $\mathcal{PT}$-Symmetric Antiferromagnets},
  author = {Ma, Da and Arora, Arpit and Vignale, Giovanni and Song, Justin C. W.},
  journal = {Phys. Rev. Lett.},
  volume = {131},
  issue = {7},
  pages = {076601},
  numpages = {6},
  year = {2023},
  month = {Aug},
  publisher = {American Physical Society},
  doi = {10.1103/PhysRevLett.131.076601},
  url = {https://link.aps.org/doi/10.1103/PhysRevLett.131.076601}
}

@article{yang2023flexible,
  title = {Flexible thermoelectric generator and energy management electronics powered by body heat},
  volume = {9},
  ISSN = {2055-7434},
  url = {http://dx.doi.org/10.1038/s41378-023-00583-3},
  DOI = {10.1038/s41378-023-00583-3},
  number = {1},
  journal = {Microsystems \& Nanoengineering},
  publisher = {Springer Science and Business Media LLC},
  author = {Yang,  Shuai and Li,  Yumei and Deng,  Ling and Tian,  Song and Yao,  Ye and Yang,  Fan and Feng,  Changlei and Dai,  Jun and Wang,  Ping and Gao,  Mingyuan},
  year = {2023},
  month = aug 
}

@article{Li2023large,
  title = {Large Anomalous Nernst Effects at Room Temperature in Fe3Pt Thin Films},
  volume = {35},
  ISSN = {1521-4095},
  url = {http://dx.doi.org/10.1002/adma.202301339},
  DOI = {10.1002/adma.202301339},
  number = {32},
  journal = {Advanced Materials},
  publisher = {Wiley},
  author = {Li,  Minghang and Pi,  Hanqi and Zhao,  Yunchi and Lin,  Ting and Zhang,  Qinghua and Hu,  Xinzhe and Xiong,  Changmin and Qiu,  Zhiyong and Wang,  Lichen and Zhang,  Ying and Cai,  Jianwang and Liu,  Wuming and Sun,  Jirong and Hu,  Fengxia and Gu,  Lin and Weng,  Hongming and Wu,  Quansheng and Wang,  Shouguo and Chen,  Yunzhong and Shen,  Baogen},
  year = {2023},
  month = jul 
}

@article{kamal2023intrinsic,
  title = {Intrinsic nonlinear conductivities induced by the quantum metric},
  author = {Das, Kamal and Lahiri, Shibalik and Atencia, Rhonald Burgos and Culcer, Dimitrie and Agarwal, Amit},
  journal = {Phys. Rev. B},
  volume = {108},
  issue = {20},
  pages = {L201405},
  numpages = {6},
  year = {2023},
  month = {Nov},
  publisher = {American Physical Society},
  doi = {10.1103/PhysRevB.108.L201405},
  url = {https://link.aps.org/doi/10.1103/PhysRevB.108.L201405}
}

@article{varshney2023quantum,
  title = {Quantum kinetic theory of nonlinear thermal current},
  author = {Varshney, Harsh and Das, Kamal and Bhalla, Pankaj and Agarwal, Amit},
  journal = {Phys. Rev. B},
  volume = {107},
  issue = {23},
  pages = {235419},
  numpages = {17},
  year = {2023},
  month = {Jun},
  publisher = {American Physical Society},
  doi = {10.1103/PhysRevB.107.235419},
  url = {https://link.aps.org/doi/10.1103/PhysRevB.107.235419}
}

@Article{yu2024ambient,
author={Yu, Ruoyao
and Feng, Shaoqing
and Sun, Qingwen
and Xu, Hao
and Jiang, Qixia
and Guo, Jinhong
and Dai, Bin
and Cui, Daxiang
and Wang, Kan},
title={Ambient energy harvesters in wearable electronics: fundamentals, methodologies, and applications},
journal={Journal of Nanobiotechnology},
year={2024},
month={Aug},
day={20},
volume={22},
number={1},
pages={497},
issn={1477-3155},
doi={10.1186/s12951-024-02774-0},
url={https://doi.org/10.1186/s12951-024-02774-0}
}

@Article{Pasquale2024,
author={Pasquale, Gabriele
and Sun, Zhe
and Migliato Marega, Guilherme
and Watanabe, Kenji
and Taniguchi, Takashi
and Kis, Andras},
title={Electrically tunable giant Nernst effect in two-dimensional van der Waals heterostructures},
journal={Nature Nanotechnology},
year={2024},
month={Jul},
day={01},
volume={19},
number={7},
pages={941-947},
issn={1748-3395},
doi={10.1038/s41565-024-01717-y},
url={https://doi.org/10.1038/s41565-024-01717-y}
}

@Article{Arisawa2024observation,
author={Arisawa, Hiroki
and Fujimoto, Yuto
and Kikkawa, Takashi
and Saitoh, Eiji},
title={Observation of nonlinear thermoelectric effect in MoGe/Y3Fe5O12},
journal={Nature Communications},
year={2024},
month={Aug},
day={26},
volume={15},
number={1},
pages={6912},
issn={2041-1723},
doi={10.1038/s41467-024-50115-4},
url={https://doi.org/10.1038/s41467-024-50115-4}
}

@article{Datta2024nonlinear,
  title = {Nonlinear Electrical Transport Unveils Fermi Surface Malleability in a Moiré Heterostructure},
  volume = {24},
  ISSN = {1530-6992},
  url = {http://dx.doi.org/10.1021/acs.nanolett.4c01946},
  DOI = {10.1021/acs.nanolett.4c01946},
  number = {31},
  journal = {Nano Letters},
  publisher = {American Chemical Society (ACS)},
  author = {Datta,  Suvronil and Bhowmik,  Saisab and Varshney,  Harsh and Watanabe,  Kenji and Taniguchi,  Takashi and Agarwal,  Amit and Chandni,  U.},
  year = {2024},
  month = jul,
  pages = {9520–9527}
}

@article{guo2024extrinsic,
  title = {Extrinsic contribution to nonlinear current induced spin polarization},
  author = {Guo, Ruda and Huang, Yue-Xin and Yang, Xiaoxin and Liu, Yi and Xiao, Cong and Yuan, Zhe},
  journal = {Phys. Rev. B},
  volume = {109},
  issue = {23},
  pages = {235413},
  numpages = {10},
  year = {2024},
  month = {Jun},
  publisher = {American Physical Society},
  doi = {10.1103/PhysRevB.109.235413},
  url = {https://link.aps.org/doi/10.1103/PhysRevB.109.235413}
}

@article{wan2025extrinsic,
  title = {Extrinsic nonlinear acoustic valley Hall effect in massive Dirac materials},
  author = {Wan, Jia-Liang and Wu, Ying-Li and Chen, Ke-Qiu and Yu, Xiao-Qin},
  journal = {Phys. Rev. B},
  volume = {111},
  issue = {4},
  pages = {045424},
  numpages = {14},
  year = {2025},
  month = {Jan},
  publisher = {American Physical Society},
  doi = {10.1103/PhysRevB.111.045424},
  url = {https://link.aps.org/doi/10.1103/PhysRevB.111.045424}
}

@article{liu2025nonlinear,
  title = {Nonlinear Nernst effect in trilayer graphene at zero magnetic field},
  ISSN = {1748-3395},
  url = {http://dx.doi.org/10.1038/s41565-025-01963-8},
  DOI = {10.1038/s41565-025-01963-8},
  journal = {Nature Nanotechnology},
  publisher = {Springer Science and Business Media LLC},
  author = {Liu,  Hao and Li,  Jingru and Zhang,  Zhifan and Zhai,  Jinfeng and Zhang,  Min and Jiang,  Hua and Xie,  X. C. and He,  Pan and Shen,  Jian},
  year = {2025},
  month = jun 
}

@article{varshney2025intrinsic,
doi = {10.1088/1367-2630/adf5dd},
url = {https://doi.org/10.1088/1367-2630/adf5dd},
year = {2025},
month = {aug},
publisher = {IOP Publishing},
volume = {27},
number = {8},
pages = {083506},
author = {Varshney, Harsh and Agarwal, Amit},
title = {Intrinsic nonlinear Nernst and Seebeck effect},
journal = {New Journal of Physics}
}

@article{hirata2025nonlinear,
  title = {Nonlinear Seebeck effect in Ni81Fe19|Pt at room temperature},
  volume = {126},
  ISSN = {1077-3118},
  url = {http://dx.doi.org/10.1063/5.0269578},
  DOI = {10.1063/5.0269578},
  number = {25},
  journal = {Applied Physics Letters},
  publisher = {AIP Publishing},
  author = {Hirata,  Y. and Kikkawa,  T. and Arisawa,  H. and Saitoh,  E.},
  year = {2025},
  month = jun 
}

@article{Ahmed2025detecting,
  title = {Detecting Lifshitz Transitions Using Nonlinear Conductivity in Bilayer Graphene},
  volume = {21},
  ISSN = {1613-6829},
  url = {http://dx.doi.org/10.1002/smll.202501426},
  DOI = {10.1002/smll.202501426},
  number = {21},
  journal = {Small},
  publisher = {Wiley},
  author = {Ahmed,  Tanweer and Varshney,  Harsh and Tu,  Bao Q. and Watanabe,  Kenji and Taniguchi,  Takashi and Gobbi,  Marco and Casanova,  Fèlix and Agarwal,  Amit and Hueso,  Luis E.},
  year = {2025},
  month = apr 
}

@misc{xiao2021lorentz,
      title={Lorentz Skew Scattering and Giant Nonreciprocal Magneto-Transport}, 
      author={Cong Xiao and Yue-Xin Huang and Shengyuan A. Yang},
      year={2025},
      eprint={2411.07746},
      archivePrefix={arXiv},
      primaryClass={cond-mat.mes-hall},
      url={https://arxiv.org/abs/2411.07746}, 
}

@article{varshney2025planar,
  title = {Planar Nernst effect from hidden band geometry in layered two-dimensional materials},
  author = {Biswas, Rahul and Varshney, Harsh and Agarwal, Amit},
  journal = {Phys. Rev. B},
  volume = {112},
  issue = {7},
  pages = {075420},
  numpages = {12},
  year = {2025},
  month = {Aug},
  publisher = {American Physical Society},
  doi = {10.1103/mqqf-fl4r},
  url = {https://link.aps.org/doi/10.1103/mqqf-fl4r}
}

@article{ma2025quantum,
  title = {Quantum kinetic theory of the semiclassical side jump, skew scattering, and longitudinal velocity},
  author = {Ma, Da and Zhang, Zhi-Fan and Jiang, Hua and Xie, X. C.},
  journal = {Phys. Rev. B},
  volume = {112},
  issue = {4},
  pages = {045136},
  numpages = {13},
  year = {2025},
  month = {Jul},
  publisher = {American Physical Society},
  doi = {10.1103/rgr5-yzy2},
  url = {https://link.aps.org/doi/10.1103/rgr5-yzy2}
}
\end{document}